\documentclass[11pt,twoside,a4paper,cmspaper,final,collab]{cms-tdr}
\def\svnVersion{dc76ac5-D}\def\svnDate{2026/10/15}\def\cmsCernNoTag{CERN-EP-2026-151}\def\cmsCernDate{\today}\def\cmsMessage{Published in the Journal of High Energy Physics as \href{http://dx.doi.org/10.1007/JHEP10(2026)084}{\doi{10.1007/JHEP10(2026)084}.}}
\begin{document}\cmsNoteHeader{HIG-25-012}

\newcommand{\PP}{\ensuremath{\Pp\Pp}\xspace}
\newcommand{\CP}{{\ensuremath{C\!P}}\xspace}
\newcommand{\phicp}{\ensuremath{\phi_{\CP}}\xspace} 
\newcommand{\alphacp}{\ensuremath{\alpha^{\PH\PGt\PGt}}\xspace} 
\newcommand{\kappattilde}{\ensuremath{\widetilde{\kappa}_\PGt}\xspace}
\newcommand{\kappat}{\ensuremath{\kappa_\PGt}\xspace}
\newcommand{\taum}{\ensuremath{\PGt_{\Pgm}}\xspace}
\newcommand{\taue}{\ensuremath{\PGt_{\Pe}}\xspace}
\newcommand{\taul}{\ensuremath{\PGt_{\Pell}}\xspace}
\newcommand{\ltau}{\ensuremath{\taul\tauh}\xspace}
\newcommand{\etau}{\ensuremath{\taue\tauh}\xspace}
\newcommand{\mutau}{\ensuremath{\PGt_{\Pgm}\tauh}\xspace}
\newcommand{\tauhtauh}{\ensuremath{\tauh\tauh}\xspace}
\newcommand{\taultaul}{\ensuremath{\taul\taul}\xspace}
\newcommand{\ggH}{\ensuremath{\Pg\Pg\PH}\xspace} 
\newcommand{\VH}{\ensuremath{\PV\PH}\xspace} 
\newcommand{\HTT}{\ensuremath{\PH\to\PGt\PGt}\xspace} 
\newcommand{\ZTT}{\ensuremath{\PZ\to\PGt\PGt}\xspace}
\newcommand{\Wjets}{\ensuremath{\PW+\text{jets}}\xspace} 
\newcommand{\Zjets}{\ensuremath{\PZ/\PGg^{*}+\text{jets}}\xspace}
\newcommand{\Zll}{\ensuremath{\PZ\to\Pell\Pell}\xspace} 
\newcommand{\Zmm}{\ensuremath{\PZ\to\Pgm\Pgm}\xspace} 
\newcommand{\Zee}{\ensuremath{\PZ\to\Pe\Pe}\xspace}
\newcommand{\jettotauh}{\ensuremath{\text{jet}\to\tauh}\xspace}
\newcommand{\ltotauh}{\ensuremath{\Pell\to\tauh}\xspace}
\newcommand{\Paone}{\ensuremath{\mathrm{a_{1}}}\xspace}
\newcommand{\PaoneONEP}{\ensuremath{\mathrm{a_{1}^{1pr}}}\xspace}
\newcommand{\PaoneTHREEP}{\ensuremath{\mathrm{a_{1}^{3pr}}}\xspace}
\newcommand{\hpm}{\ensuremath{\text{h}^\pm}\xspace}
\newcommand{\hmp}{\ensuremath{\text{h}^\mp}\xspace}
\newcommand{\rhorho}{\ensuremath{\PGr\PGr}\xspace}
\newcommand{\rhoaone}{\ensuremath{\PGr\PaoneTHREEP}\xspace}  
\newcommand{\murho}{\ensuremath{\Pgm\PGr}\xspace} 
\newcommand{\erho}{\ensuremath{\Pe\PGr}\xspace} 
\newcommand{\pirho}{\ensuremath{\Pgp\PGr}\xspace}
\newcommand{\dz}{\ensuremath{d_{z}}\xspace}
\newcommand{\dxy}{\ensuremath{d_{xy}}\xspace}
\newcommand{\SIP}{\ensuremath{S_{\text{IP}}}\xspace}
\newcommand{\Deltaeta}{\ensuremath{\Delta\eta}\xspace}
\newcommand{\Deltaphi}{\ensuremath{\Delta\phi}\xspace}
\newcommand{\FASTMTT}{{\textsc{FastMTT}}\xspace}
\newcommand{\FF}{\ensuremath{F_{\text{F}}}\xspace} 
\newcommand{\TAUSPINNER}{{\textsc{tauspinner}}\xspace}
\newcommand{\thetaGJ}{\ensuremath{\theta_{\mathrm{GJ}}}\xspace}
\newcommand{\thetaGJmax}{\ensuremath{\theta_{\mathrm{GJ}}^{\text{max}}}\xspace}
\newcommand{\mtau}{\ensuremath{m_{\PGt}}\xspace}
\newcommand{\maone}{\ensuremath{m_{\Paone}}\xspace}
\newcommand{\paone}{\ensuremath{\vec{p}_{\Paone}}\xspace}
\newcommand{\vectheta}{\ensuremath{\vec{\theta}}\xspace}
\newcommand{\likeli}{\ensuremath{L(\muggh, \muqqh,\alphacp,\vec{\theta})}\xspace}
\newcommand{\muqqh}{\ensuremath{\mu_{\Pq\Pq\PH}}\xspace} 
\newcommand{\muggh}{\ensuremath{\mu_{\Pg\Pg\PH}}\xspace}
\newcommand{\nij}{\ensuremath{n_{i}}\xspace} 
\newcommand{\Bi}{\ensuremath{B_{i}(\vectheta)}\xspace} 
\newcommand{\Si}{\ensuremath{S_{i}(\muggh,\muqqh,\alphacp,\vectheta)}\xspace} 
\newcommand{\Cm}{\ensuremath{C_{m} (\vectheta)}\xspace} 
\newcommand{\AveAsym}{\ensuremath{A\:S/(S+B)}\xspace}

\cmsNoteHeader{HIG-25-012}

\title{Analysis of the \texorpdfstring{\CP}{CP} structure of the Yukawa coupling between the Higgs boson and tau leptons in proton-proton collisions at \texorpdfstring{$\sqrt{s}=13.6\TeV$}{sqrt(s) = 13.6 TeV}}

\hypersetup{
    pdftitle={Analysis of the CP structure of the Yukawa coupling between the Higgs boson and tau leptons in proton-proton collisions at 13.6 TeV},
    pdfauthor={CMS Collaboration},
    pdfsubject={CMS},
    pdfkeywords={CMS,  Higgs, taus, CP}
}

\date{\today}

\abstract{
 This paper presents a measurement of the charge-parity (\CP) structure of the Yukawa coupling between the Higgs boson (\PH) and tau leptons, using proton-proton collision data at $\sqrt{s}=13.6\TeV$ recorded with the CMS detector at the LHC, corresponding to an integrated luminosity of 62.4\fbinv. Angular correlations between the decay products of tau leptons produced in $\HTT$ decays are exploited to constrain the effective \CP mixing angle $\alphacp$, which parameterizes the admixture of scalar and pseudoscalar couplings. The mixing angle is measured to be $\alphacp = (36^{+33}_{-30})^\circ$, compared with an expected value of $(0 \pm 19)^\circ$ under the standard model hypothesis. When combined with the previous CMS measurement using data collected at $\sqrt{s} = 13\TeV$, corresponding to an integrated luminosity of 138\fbinv, the mixing angle is determined to be $(7 \pm 16)^\circ$, with an expected value of $(0 \pm 14)^\circ$. This result represents the most precise measurement by CMS of the \CP nature of the Higgs boson coupling to tau leptons, with an expected precision that is the best achieved by any experiment to date.
}

\hypersetup{
pdfauthor={L. Russell, D. Winterbottom, D. Colling (CCLE)},
pdftitle={Analysis of the CP Structure of the Yukawa Coupling between the Higgs boson and tau leptons},
pdfsubject={CMS},
pdfkeywords={CMS, your topics}}

\maketitle 

\section{Introduction}

In the standard model (SM) electroweak (EW) symmetry breaking is postulated via the Brout--Englert--Higgs mechanism~\cite{PhysRevLett.13.321, Higgs:1964ia, PhysRevLett.13.508, PhysRevLett.13.585, Higgs:1966ev, Kibble:1967sv}. 
This mechanism predicts the existence of a scalar particle, the Higgs boson (\PH).  
A particle compatible with this boson was discovered by the ATLAS~\cite{Aad_2012} and CMS~\cite{Chatrchyan_2012, Chatrchyan:2013lba} Collaborations at the CERN LHC in 2012.
Since then, the couplings of the Higgs boson to heavy quarks, leptons, and gauge bosons have been measured and found to be consistent with the SM predictions within the current experimental uncertainties~\cite{CMS:Combination,ATLAS:Combination}.

Despite the many successes of the SM, it is known to be incomplete as it does not account for all observed phenomena. 
For instance, it does not provide an explanation for the observed baryon asymmetry of the universe (BAU).
One of the necessary conditions to generate this asymmetry is the violation of the charge-parity (\CP) symmetry~\cite{Sakharov:1967dj}. 
While \CP violation is present in the quark sector of the SM, the amount is insufficient to explain the observed BAU by many orders of magnitude~\cite{Gavela:1994dt}. 
Additional sources of \CP violation are therefore required.

An intriguing possibility is that new sources of \CP violation may arise from the Higgs sector.
The SM assumes a minimal Higgs sector, which predicts a single \CP-even Higgs boson.
However, nonminimal Higgs sectors can introduce additional sources of \CP violation. In such models the Higgs boson can be a mixture of \CP-even and \CP-odd states, leading to \CP-violating interactions.
For example, the complex two-Higgs-doublet model~\cite{Biekotter:2024ykp} and the next-to-minimal supersymmetric SM~\cite{King_2015} can incorporate \CP violation in the Higgs sector.

The \CP properties of Higgs boson interactions with vector bosons have been studied extensively by the ATLAS and CMS Collaborations~\cite{Chatrchyan:2012jja,Chatrchyan:2013mxa,Khachatryan:2014kca,Khachatryan:2016tnr,Sirunyan:2017tqd,Sirunyan_2019,CMS:2021nnc,Aad:2013xqa,Aad:2015mxa,Aad:2016nal,Aaboud:2017vzb,Aaboud:2018xdt}. 
The measurements of interactions with the \PW and \PZ bosons (collectively referred to as \PV bosons) have excluded the possibility that the Higgs boson is a pure pseudoscalar (\CP-odd).
However, these measurements do not rule out the Higgs boson having a mixed \CP state, with both \CP-even and \CP-odd components.
In fact, the bounds on mixed \CP states are relatively weak, as there is no tree-level \CP-odd coupling between the Higgs boson and the \PV bosons.
Any \CP-odd contribution to the \PH-\PV boson interactions can only be generated through loop processes, and is, therefore, suppressed compared with the \CP-even coupling.

The couplings of the Higgs boson to fermions can have both \CP-even and \CP-odd components at tree level, which makes them particularly interesting for detecting \CP violation.
The experimentally accessible fermionic \CP measurements at the LHC are those involving the top quark and the \PGt lepton Yukawa couplings~\cite{Gritsan_2016}.\footnote{The $\PH\to\bbbar$ channel is less suitable for probing the \CP structure of the
Yukawa coupling because bottom quarks are observed only as hadronic jets, and the necessary spin information of the original quarks is largely washed out by the
hadronization process. 
Although heavy-quark spin information can in principle be accessed via the polarization of baryons produced in their fragmentation,
such as $\Lambda_{\mathrm{b}}$, the small fragmentation fractions, decay branching fractions, and challenging reconstruction lead to a severe loss of
efficiency, making this approach difficult to exploit for Higgs boson \CP studies~\cite{Kats:2023zxb}.}
The ATLAS and CMS Collaborations have studied the \CP structure of the top quark Yukawa coupling using gluon-gluon fusion (\ggH)~\cite{Aaboud:2018xdt,CMS:2021nnc,CMS:2022uox,CMS:2024bua,ATLAS:2025hki} and associated production with a top quark and antiquark pair~\cite{CMSTOPCP2020,ATLASTOPCP2020,CMS:2021nnc,CMS:2022dbt,ATLAS:2023cbt,CMS:2024fdo,ATLAS:2025eua}, excluding large \CP-odd contributions.

While the top coupling can, {\it{a priori}}, contain a sufficiently large \CP-odd component to account for the BAU, it is tightly constrained by indirect measurements. Notably stringent limits on electric dipole moments, particularly that of the electron, have been set~\cite{Roussy:2022cmp}.
These constraints restrict the viable parameter space, disfavouring explanations of the BAU based solely on \CP violation in the top quark Yukawa coupling.
In contrast, the \PGt lepton Yukawa coupling is less constrained by indirect measurements, and its \CP-odd component can be sufficiently large to account for the BAU while remaining compatible with the experimental bounds~\cite{bahl_2022}.
The first direct measurement of the \CP structure of the Higgs boson coupling to \PGt leptons was presented by the CMS Collaboration in Ref.~\cite{run2_cp_paper}, and was later followed by a measurement by the ATLAS Collaboration in Ref.~\cite{ATLAS:2022akr}.
Both analyses excluded the pure \CP-odd coupling hypothesis at significance levels exceeding three standard deviations.
Mixed states with large \CP-odd contributions are also disfavoured, although a sizeable \CP-odd component remains allowed within the current experimental precision.

In this paper we present a new measurement of the \CP structure of the Higgs boson coupling to \PGt leptons.
This analysis uses a proton-proton (\PP) data set collected with the CMS detector at $\sqrt{s}=13.6\TeV$ in 2022 and 2023, corresponding to an integrated luminosity of 62.4\fbinv. 
When combined with the previous CMS measurement, this analysis provides the most precise measurement by CMS of the \CP nature of the \Pgt Yukawa coupling,
with an expected precision that is the best achieved by any experiment to date.

This paper is organized as follows. 
Section~\ref{sec:theory} introduces the parameterization of the \CP structure of the H-\PGt lepton Yukawa coupling.
Section~\ref{sec:CMSdetector} outlines the CMS detector, which is followed by a description of the event reconstruction in Section~\ref{sec:eventreconstruction}.
The \CP-sensitive observables are outlined in Section~\ref{sec:decayplanereco}, and an overview of the simulated samples used in the analysis is given in Section~\ref{sec:datasimulatedsamples}.
The event selections are described in Section~\ref{sec:eventselection}, and the estimation of the backgrounds is discussed in Section~\ref{sec:backgroundestimation}. 
The techniques used to distinguish the signal from the background events are outlined in Section~\ref{sec:eventcategorisation}. 
Section~\ref{sec:unrolled} displays various distributions that are used to extract the results, and the systematic uncertainties are presented in Section~\ref{sec:systematicuncertainties}. 
The results, along with a projection of the measurement to the end of the High-Luminosity LHC, are discussed in Section~\ref{sec:results}, and a summary of the paper is given in Section~\ref{sec:summary}.
Tabulated results are provided in the HEPData record for this analysis~\cite{hepdata}.

\section{Parameterization of the \texorpdfstring{\CP structure of the $\PH\PGt\PGt$}{CP structure of the Htautau} coupling}
\label{sec:theory}

We express the Lagrangian for the \PGt lepton Yukawa coupling in terms of the coupling strength modifiers \kappat and \kappattilde, which parameterize the \CP-even and \CP-odd components of the interaction, respectively~\cite{Gritsan_2016}: 
\begin{linenomath}
\begin{equation}
\mathcal{L}_{\mathrm{Y}}=-\frac{\mtau}{v}\PH(\kappat\PAGt\PGt+\kappattilde\PAGt i\gamma_{5}\PGt),
\end{equation}
\end{linenomath}
where \mtau is the \PGt lepton mass, \PGt denotes the Dirac spinor of \PGt lepton fields, and the vacuum expectation value of the Higgs field, $v$, has a value of 246\GeV. In the SM, the \PGt lepton Yukawa coupling is purely \CP-even, which corresponds to $\kappat=1$ and $\kappattilde=0$.

The \CP properties of the \HTT process are commonly described in terms of an effective mixing angle \alphacp, defined as 
\begin{linenomath}
\begin{equation}
\tan(\alphacp)=\frac{\kappattilde}{\kappat}.
\label{eq:phicpinkappa}
\end{equation}
\end{linenomath}
A mixing angle of $\alphacp=0^{\circ}\,(90^{\circ})$ corresponds to a pure scalar (pseudoscalar) coupling. For any other value of \alphacp, the Higgs boson has a \CP-violating coupling.

\section{The CMS detector}\label{sec:CMSdetector}

The CMS apparatus~\cite{CMS:2008xjf,CMS:2023gfb} is a multipurpose, nearly hermetic detector, designed to trigger on~\cite{CMS:2020cmk,CMS:2016ngn,CMS:2024aqx} and identify electrons, muons, photons, and (charged and neutral) hadrons~\cite{CMS:2020uim,CMS:2018rym,CMS:2014pgm}. 
Its central feature is a superconducting solenoid of 6\unit{m} internal diameter, providing a magnetic field of 3.8\unit{T}. 
Within the solenoid volume are a silicon pixel and strip tracker, a lead tungstate crystal electromagnetic calorimeter (ECAL), and a brass and scintillator hadron calorimeter (HCAL), each composed of a barrel and two endcap sections. 
Forward calorimeters extend the pseudorapidity ($\eta$) coverage provided by the barrel and endcap detectors. 
Muons are reconstructed using gas-ionization detectors interleaved with the layers of the steel flux-return yoke outside the solenoid.
A more detailed description of the CMS detector, together with a definition of the coordinate system used and the relevant kinematic variables, can be found in Refs.~\cite{CMS:2008xjf,CMS:2023gfb}.

Events of interest are selected using a two-tiered trigger system. 
The first level, composed of custom hardware processors, uses information from the calorimeters and muon detectors to select events at a rate of around 100\unit{kHz} within a fixed latency of 4\mus~\cite{CMS:2020cmk}. 
The second level, known as the high-level trigger, consists of a farm of processors running a version of the full event reconstruction software optimized for fast processing, and reduces the event rate to a few kHz before data storage~\cite{CMS:2016ngn,CMS:2024aqx}.

\section{Event reconstruction} 
\label{sec:eventreconstruction}

A particle-flow (PF) algorithm~\cite{CMS:2017yfk} aims to reconstruct and identify each individual particle in an event, with an optimized combination of information from the various elements of the CMS detector. 
The energy of photons is obtained from the ECAL measurement. 
The energy of an electron is determined from a combination of the electron momentum at the primary interaction vertex (PV) as determined by the tracker, the energy of the corresponding ECAL cluster, and the energy sum of all bremsstrahlung photons spatially compatible with originating from the electron track. 
The energy of a muon is obtained from the curvature of the corresponding track. The energy of a charged hadron is determined from a combination of the momentum measured in the tracker and the matching ECAL and HCAL energy deposits, corrected for the response function of the calorimeters to hadronic showers. 
Finally, the energy of neutral hadrons is obtained from the corresponding corrected ECAL and HCAL energies. 

\subsection{Primary vertex}

The positions of all \PP interactions (vertices) in the event, including the hard scatter (PV) and soft (pileup) vertices, are reconstructed in a two-step procedure~\cite{CMS:2014pgm}. 
The first step clusters the tracks that appear to originate from the same interaction using the deterministic annealing algorithm~\cite{Rose:1998}. 
The second step fits the position of each vertex using tracks associated to its cluster with the adaptive vertex fitter (AVF) algorithm~\cite{Waltenberger_2007}. 
The PV is taken to be the vertex corresponding to the hardest scattering in the event, evaluated using tracking information alone, as described in Section 9.4.1 of Ref.~\cite{CMS-TDR-15-02}.

The LHC beamspot represents a three-dimensional profile of the luminous region, where the LHC beams collide in the CMS detector. 
The parameters of the beamspot are determined from an average over many events~\cite{CMS:2014pgm}. 
The uncertainties in the beamspot parameters are relatively small and are incorporated into the AVF algorithm to provide an additional constraint on the PV position.
The inclusion of the beamspot information leads to an improvement of the PV resolution in the transverse plane, while the $z$ coordinate of the PV is largely unaffected.
We will refer to this estimate of the \PP interaction point as the ``refitted PV''.

\subsection{Jets and missing transverse momentum}

For each event, hadronic jets are clustered from reconstructed PF particles using the infrared and collinear safe anti-\kt algorithm~\cite{Cacciari:2008gp, Cacciari:2011ma} with a distance parameter of 0.4.
Jet momentum is determined as the vectorial sum of all particle momenta in the jet, and is found from simulation to be, on average, within 5 to 10\% of the true momentum over the whole transverse momentum (\pt) spectrum and detector acceptance. 
Additional pp interactions within the same or nearby bunch crossings (pileup) can contribute extra tracks and calorimetric energy depositions, increasing the apparent jet momentum. 
The pileup per particle identification algorithm (PUPPI)~\cite{Sirunyan:2020foa,Bertolini:2014bba} is used to mitigate the effect of pileup at the reconstructed-particle level, making use of local shape information, event pileup properties, and tracking information. 

Jet energy corrections are derived from simulation studies so that the average measured energy of jets becomes identical to that of particle level jets. 
In situ measurements of the momentum balance in dijet, $\text{photon} + \text{jet}$, $\PZ + \text{jet}$, and multijet events are used to determine any residual differences between the jet energy scale in data and in simulation, and appropriate corrections are made~\cite{CMS:2016lmd}. 
Additional selection criteria are applied to each jet to remove jets potentially dominated by instrumental effects or reconstruction failures~\cite{Sirunyan:2020foa}. 

The missing transverse momentum vector \ptvecmiss is computed as the negative vector sum of the \pt of all the PF candidates in an event, and its magnitude is denoted as \ptmiss~\cite{CMS:2019ctu}. 
The \ptvecmiss is modified to account for corrections to the energy scale of the reconstructed jets in the event.
The PUPPI algorithm is applied to reduce the pileup dependence of the \ptvecmiss observable~\cite{CMS:2019ctu}. 

\subsection{Muons}

Muons are measured in the range $\abs{\eta} < 2.4$, with detection planes made using three technologies: drift tubes, cathode strip chambers, and resistive plate chambers. 
The single muon trigger efficiency exceeds 90\% over the full $\eta$ range, and the efficiency to reconstruct and identify muons is greater than 96\%. 
Matching muons to tracks measured in the silicon tracker results in a relative transverse momentum resolution, for muons with \pt up to 100\GeV, of 1\% in the barrel and 3\% in the endcaps. 
Measurements made with cosmic ray muons show that, in the central region of the detector, the \pt resolution is better than 7\% for muons with \pt up to 1\TeV~\cite{CMS:2018rym}.

In order to reject muons that originate from nonprompt interactions, or are misidentified, muons are required to have a relative isolation~\cite{CMS:2018rym} of less than $0.15$. 
Muon candidates used to tag \PGt lepton decays into muons (\taum) are required to have $\pt > 26\GeV$ and $\abs{\eta} < 2.4$.

\subsection{Electrons}

Electrons are reconstructed in the range $\abs{\eta} < 2.5$ by combining tracks and ECAL energy clusters.
Their identification is based on a boosted decision tree (BDT) algorithm sensitive to the topology of the energy deposits in the ECAL, the quality of tracks, the compatibility of the momentum measurement from the tracker and the energy measured by the ECAL, and their isolation~\cite{CMS:2020uim}.

The electron momentum is estimated by combining the energy measurement in the ECAL with the momentum measurement in the tracker. 
The momentum resolution for electrons with $\pt \approx 45\GeV$ from $\PZ \to \Pe \Pe$ decays ranges from 1.6 to 5\%. 
It is generally better in the barrel region than in the endcaps, and also depends on the bremsstrahlung energy emitted by the electron as it traverses the material in front of the ECAL~\cite{CMS:2020uim,CMS-DP-2020-021}. 

Electron candidates used to tag \PGt lepton decays into electrons (\taue) are required to have $\pt > 32\GeV$ and $\abs{\eta} < 2.1$.

\subsection{Hadronic taus}

Hadronically decaying \PGt leptons (\tauh) are reconstructed from jets, using the hadrons-plus-strips (HPS) algorithm~\cite{CMS:2018jrd}, which combines one or three tracks with energy deposits in the calorimeters.
Neutral pions (\Pgpz) are reconstructed as strips with dynamic size in $\eta$-$\phi$ from reconstructed photons, and electrons from photon conversions, where $\phi$ is the azimuthal angle in radians and the strip size varies as a function of the $\pt$ of the electron or photon candidate.
The energy of the \Pgpz is estimated by summing the energies of all electron/photon candidates collected by the HPS algorithm. 
The direction of the \Pgpz is taken as the direction of the leading electron/photon candidate.
In most cases the leading candidate is a photon and the direction is determined by pointing its associated ECAL clusters back towards the PV.
Finally, the mass is set to the known \Pgpz mass.

To distinguish genuine \tauh decays from jets originating from the hadronization of quarks or gluons, and from electrons, or muons, the \textsc{DeepTau} algorithm is used~\cite{CMS:2022prd,CMS:2025kgf}. 
Information from all individual reconstructed particles near the \tauh axis is combined with properties of the $\tauh$ candidate and the event.

The HPS algorithm, combined with \textsc{DeepTau}, achieves an efficiency of about 45\% for genuine \tauh candidates with $\pt > 20\GeV$, while reducing the probability for jets to be misidentified as \tauh to $\mathcal{O}(10^{-3})$~\cite{CMS:2022prd,CMS:2025kgf}.

Table~\ref{tab:AnalysisStrategy_TauDecayModes} summarizes the \PGt lepton decay modes used in this analysis, their branching fractions, and the shorthand symbols that we use to denote them in the rest of this paper. 
The charged hadrons are denoted by the symbol \hpm, which consist mainly of charged pions but include a smaller contribution from charged kaons. Throughout this paper we will assume that all \hpm are charged pions (\Pgppm) as the CMS detector is not able to distinguish between pions and kaons at this momentum scale.

\begin{table}[hbtp]
\centering
\topcaption{Decay modes of \PGt leptons used in this analysis and their branching fractions $\mathcal{B}$~\cite{PDG2024}. Where appropriate, we indicate the known intermediate resonances. The last row gives the shorthand notation for the decays used throughout this paper. \label{tab:AnalysisStrategy_TauDecayModes}}
\begin{tabular}{cccccccc}
\hline
\noalign{\vskip 2.5pt}
Mode & $\Pepm\PGn\PGn$ & $\PGmpm\PGn\PGn$ & $\hpm\nu$ & $\hpm\Pgpz\nu$ & $\hpm\Pgpz\Pgpz\PGn$  & $\hpm\hmp\hpm\nu$ \\ \hline
Type &\taue        &\taum           &\tauh   &\tauh  & \tauh  &\tauh \\
$\mathcal{B}(\%)$& 17.8 & 17.4 & 11.5 & 25.9 & 9.5 & 9.8 \\
Resonance & \NA & \NA & \NA & \Pgr & \Pai  & \Pai \\
Symbol &\Pe        &\Pgm           &\Pgp   &\PGr  & \PaoneONEP  &\PaoneTHREEP \\ \hline
\end{tabular}
\end{table}

The \tauh candidate decay modes are determined using the \textsc{ParticleNet} algorithm~\cite{PhysRevD.101.056019}, an architecture based on a graph neural network exploiting properties of the reconstructed particles in the anti-\kt jets.
This method achieves a purity of approximately 80\% for the \Pgp and \PGr, 70\% for the \PaoneONEP and 90\% for the \PaoneTHREEP topologies. This corresponds to an improvement of approximately 10\% across all decay modes compared with previous methods~\cite{CMS-DP-2020-041}.
Leptonic \PGt decays (\taul) are reconstructed as muons (for \taum) or electrons (for \taue).
The \tauh candidates are required to have $\pt > 20\GeV$ and $\abs{\eta} < 2.5$.

\section{Reconstruction of \texorpdfstring{\CP}{CP}-sensitive observables}\label{sec:decayplanereco}

The \CP nature of the \HTT interaction affects the spin correlations between the two \PGt leptons produced in the decay. 
This, in turn, modifies the angular distributions of the \PGt lepton decay products, which can be exploited to measure \alphacp.

\begin{figure}[!htb]
    \centering
    \includegraphics[width=0.45\textwidth]{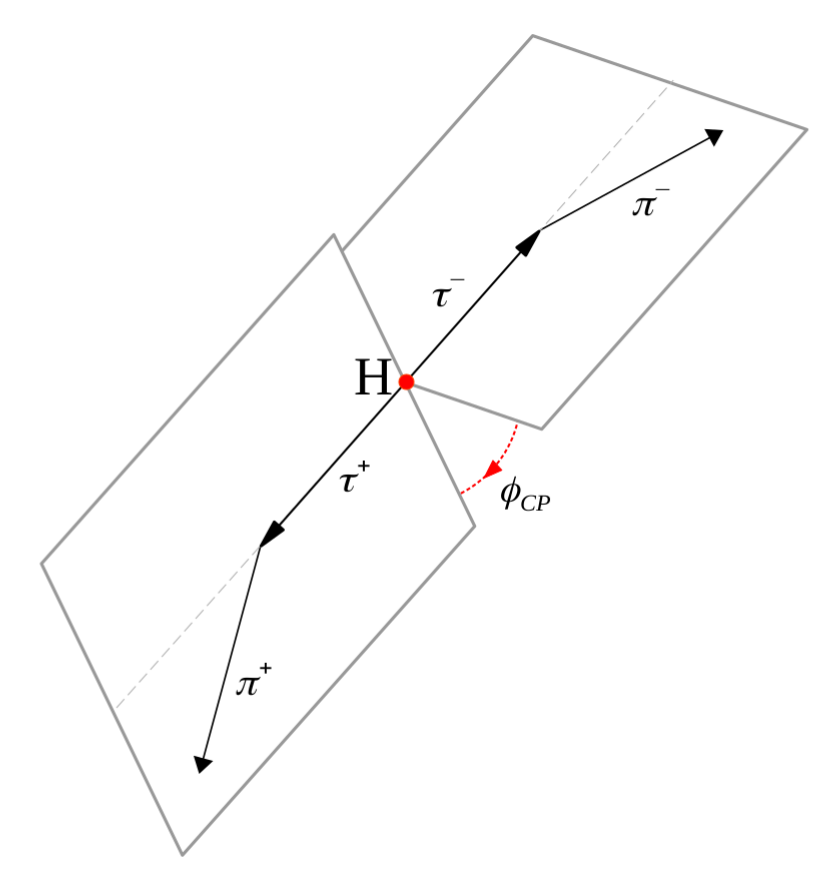}
    \caption{The decay planes of two \PGt leptons each decaying to a charged pion and a neutrino. The angle \phicp is defined as the angle between the two decay planes in the Higgs boson rest frame.}
    \label{fig:decay_planes}
\end{figure}

The \CP-mixing angle \alphacp is determined from the acoplanarity angle \phicp, defined as the angle between the \PGt lepton decay planes in the Higgs boson rest frame, which is sensitive to the \CP structure of the \PGt Yukawa coupling.
An illustration of the decay planes for the final state in which both \PGt leptons decay as $\PGt^\mp\rightarrow\pi^\mp\nu$ is shown in Fig.~\ref{fig:decay_planes}. 

The relation between \alphacp and \phicp for the decay of a Higgs boson via \PGt leptons to two outgoing charged particles~\cite{BERGE2016841} is:
\begin{linenomath}
\begin{equation}
\dd{\Gamma}{\phicp}(\PH\to \PGtp\PGtm)\sim 1-b(E^{+})b(E^{-})\frac{\pi^2}{16}\cos(\phicp-2\alphacp).
\label{eq:DifTauDecayWidth}
\end{equation}
\end{linenomath}
In this equation, the outgoing charged particles have an energy $E^{\pm}$ in their respective \PGt lepton rest frames. The correlation between the momentum of the charged particle and the \PGt spin is described by spectral functions~\cite{Berge:2011ij}, denoted $b$, which are different for each \PGt lepton decay mode.
In Higgs boson decays to a pair of \PGt leptons, there is, therefore, a phase shift between different \CP-mixing scenarios such that the difference in \phicp equals two times the difference in \alphacp, as given by Eq.~(\ref{eq:DifTauDecayWidth}). The background processes (including Drell--Yan) do not exhibit such a phase shift.

In practice, the reconstruction of \phicp is complicated by the fact that the \PGt lepton momenta cannot be fully reconstructed, owing to the presence of neutrinos in the final state. Approximate methods must therefore be used to define the decay planes and extract the \CP-sensitive variables, which for convenience we continue to refer to collectively as \phicp.
Various approaches to reconstruct \phicp have been proposed in the literature~\cite{Kramer:1993jn,Bower:2002zx,Desch:2003mw,Desch:2003rw,Berge:2008wi,Berge:2008dr,Berge:2011ij,Berge:2013jra,Berge:2014sra,Berge:2015nua,Jozefowicz:2016kvz,Cherepanov:2018yqb}.
These can broadly be categorised into three classes: (i) the \textit{impact parameter method}, which uses the impact parameter vector of the charged decay product in place of the unmeasured \PGt lepton direction; (ii) the \textit{neutral pion method}, which replaces the \PGt lepton decay plane with the decay plane of the intermediate $\rho$ resonance in the $\PGt^{\mp}\rightarrow\rho^\mp(\rightarrow\pi^\mp\pi^0)\nu$ decay; and (iii) the \textit{polarimetric vector method}, which exploits \PGt lepton decays involving three charged particles, using the secondary decay vertex to constrain the \PGt lepton momentum and the reconstructed \PGt lepton polarimetric vector to define the plane.

In this paper, we use a combination of these methods to reconstruct \phicp for the various \PGt lepton decay modes.
For final states not involving \PaoneTHREEP decays, we follow the definitions of \phicp given in Ref.~\cite{Berge:2015nua}, based on the impact parameter and neutral pion methods, whereas for the $\PaoneTHREEP\PaoneTHREEP$ channel we use the polarimetric vector method, as defined in Ref.~\cite{Cherepanov:2018yqb}.
For $\PaoneTHREEP+X$ final states, where $X\neq\PaoneTHREEP$, we define a hybrid approach combining these two methods.
The explicit definitions and reconstruction procedures for \phicp used in this work are described below.

\subsection{Definition of vectors}

For each $\PGt^\pm$ lepton decay, we define vectors  $\vec{\textrm{P}}^\pm$ and $\vec{\textrm{R}}^\pm$, which are used to construct the decay planes of the \PGt leptons (a similar notation is adopted in Ref.~\cite{Cardini:2022qqy}).
If the \PGt decays via the $\pi$, \Pe or \PGm modes, the vector $\vec{\textrm{P}}^\pm$ is the momentum of the charged decay product (the \PGp, \PGm, or \Pe).
The vector $\vec{\textrm{R}}^\pm$ is defined as the vector between the refitted PV and the point of closest approach of the charged particle's track to this vertex. The magnitude of $\vec{\textrm{R}}^\pm$ gives the impact parameter of the charged decay product.

In cases where the \PGt decays via the \PGr or \PaoneONEP mode, the vector $\vec{\textrm{P}}^\pm$ is also the momentum of the charged pion (\PGp). However, the vector $\vec{\textrm{R}}^\pm$ is instead defined as the momentum of the neutral component, which corresponds to a single \Pgpz in the \PGr mode, or the sum of two \Pgpz~s in the \PaoneONEP mode.  

Finally, for the \PaoneTHREEP mode, the full \PGt lepton momentum can be well reconstructed, and is used to define $\vec{\textrm{P}}^\pm$.
The direction of the \PGt momentum is taken as the vector between the secondary \PGt decay vertex and the refitted PV. If both \PGt leptons in the pair decay via the \PaoneTHREEP mode, the magnitude of each \PGt lepton momentum is determined using~\cite{Cherepanov:2018npf}
\begin{linenomath}
\begin{equation}
\abs{\vec{p}_{\PGt}}=\frac{(\maone^2+\mtau^2)\abs{\paone}\cos\thetaGJ\pm\sqrt{\smash[b]{(\maone^2 + \abs{\paone}^2)( (\maone^2-\mtau^2)^2 -4\mtau^2\abs{\paone}^2 \sin^2 \thetaGJ )}}}{2(\maone^2+\abs{\paone}^2\sin^2 \thetaGJ)},
\label{eq:pola_vec}
\end{equation}
\end{linenomath}
where the Gottfried--Jackson angle \thetaGJ is defined as the angle between the intermediate \Paone momentum and the \PGt lepton momentum in the laboratory frame~\cite{Cherepanov:2018npf}. 
The maximally allowed value of \thetaGJ, denoted \thetaGJmax, is given by $\arcsin((\mtau^2-\maone^2)/2\maone\abs{\paone})$.
For decays in which the reconstructed \thetaGJ exceeds \thetaGJmax, due to the finite angular resolution of the charged pions and \PGt direction measurements, the value of \thetaGJ is set to \thetaGJmax. The ambiguity in the analytical solution above is resolved by selecting the pair of solutions with the mass closest to that of the Higgs boson.
If only one \PGt lepton decays via the \PaoneTHREEP mode, the magnitude of the \PGt momentum is instead estimated using the \FASTMTT algorithm~\cite{fastmtt_2025}, with the mass of the ditau pair constrained to a narrow range around the Higgs boson mass.
The $\vec{\textrm{R}}^\pm$ vector for \PaoneTHREEP decays is the polarimetric vector, which is an estimate of the most likely direction of the \PGt spin~\cite{Davier:1992nw}. The polarimetric vector is defined using the \PGt lepton and \PaoneTHREEP (visible \PGt) momenta~\cite{Cherepanov:2023wfp}, using the \PaoneTHREEP resonance model as implemented in the \TAUOLA~\cite{Jadach:1990mz,Jezabek:1991qp,Jadach:1993hs} program, which uses parameters measured by the CLEO Collaboration~\cite{Asner:1999kj}. 

\subsection{Determining the acoplanarity angle}

\begin{figure}[htb!]
\centering
\includegraphics[width=0.95\textwidth]{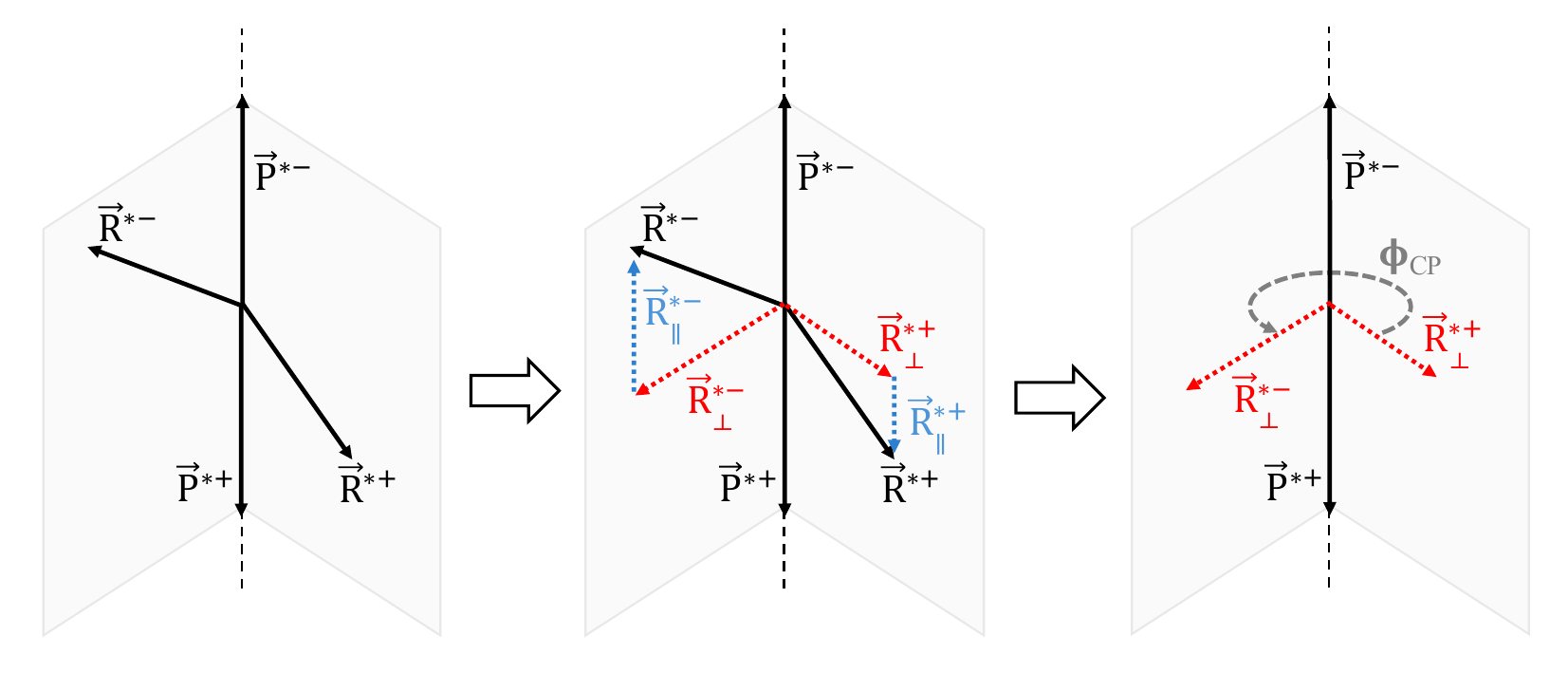}
\caption{Illustration of the reconstruction of \phicp. The frame is defined such that the sum of the $\vec{\textrm{P}}^{*\pm}$ vectors is zero. The \phicp angle is reconstructed from $\vec{\textrm{R}}^{*\pm}_\perp$ and $\vec{\textrm{P}}^{*\pm}$.
\label{fig:decayplanes}}
\end{figure}

The $\vec{\textrm{R}}^\pm$ and $\vec{\textrm{P}}^\pm$ vectors are boosted into the frame where the $\vec{\textrm{P}}^\pm$ vectors sum to zero, these boosted vectors are denoted $\vec{\textrm{R}}^{*\pm}$ and $\vec{\textrm{P}}^{*\pm}$. Each boosted $\vec{\textrm{R}}^{*\pm}$ vector is then split into components parallel and perpendicular to the $\vec{\textrm{P}}^{*\pm}$ vectors. The perpendicular components $\vec{\textrm{R}}^{*\pm}_\perp$ are then used to determine the angle between planes, as illustrated in Fig.~\ref{fig:decayplanes}.

The corresponding unit vectors are used to define the angle $\phi^*$:
\begin{equation}
	\phi^* = \textrm{arccos}(\hat{\textrm{R}}^{*+}_\perp \cdot\hat{\textrm{R}}^{*-}_\perp),
\end{equation}
and the \CP-odd triple product correlation $O^{*}$ observable:
\begin{equation}
	O^{*} = \hat{\textrm{P}}^{*-}\cdot (\hat{\textrm{R}}^{*+}_\perp \times \hat{\textrm{R}}^{*-}_\perp).
\end{equation}

The acoplanarity angle \phicp is then defined as:
\begin{equation}
\phicp =
\begin{cases}
\phi^{*}, & \text{if } O^{*} \ge 0, \\
2\pi - \phi^{*}, & \text{if } O^{*} < 0.
\end{cases}
\end{equation}

In configurations involving \PGr or \PaoneONEP modes an additional shift is required to resolve \phicp (due to the different possible polarizations of the intermediate mesons). The quantity
\begin{equation}
	y^{\PGt\pm} = \frac{\textrm{E}(\Pgpz) - \textrm{E}(\PGp^\pm)}{\textrm{E}(\Pgpz) + \textrm{E}(\PGp^\pm)}
\end{equation}
is calculated for each decay involving a \PGr or an \PaoneONEP, and used to separate the different polarizations.
For channels involving one (or two) such decay(s), if $y^{\PGt\pm}$ (or $y^{\PGt+} y^{\PGt-}$) is negative, a transformation is applied to \phicp:
\begin{equation}
\phi_{CP}^{\text{New}} =
\begin{cases}
\phicp + \pi, & \text{if } \phicp < \pi, \\
\phicp - \pi, & \text{if } \phicp \ge \pi.
\end{cases}
\end{equation}

\section{Data and simulated event samples}\label{sec:datasimulatedsamples}

The data sample used in this analysis corresponds to an integrated luminosity of 62.4\fbinv collected at $\sqrt{s}=13.6\TeV$. 
The signal and relevant background processes are modelled with samples of Monte Carlo (MC) simulated events.

The signal samples with a Higgs boson produced through \ggH, vector boson fusion (VBF), or in association with a \PW or \PZ vector boson (denoted as $\PW\PH$ or $\PZ\PH$, or \VH when combined) are generated at next-to-leading order (NLO) in perturbative quantum chromodynamics (QCD) with the \POWHEG 2.0~\cite{Nason:2004rx,Frixione:2007vw, Alioli:2010xd, Nason:2009ai, Jezo:2015aia, Granata:2017iod, Nason:2020lxx} event generator. 
The Higgs boson production mechanism is configured to only produce scalar Higgs bosons, since in our analysis of \alphacp we are not sensitive to the modifications to the \CP-nature of the production interactions. 
The \ggH simulation does not account for quark mass effects in the quark loop, which mainly affect the high-\pt region.
Therefore, the Higgs boson \pt spectrum in the \ggH sample is reweighted by applying \pt-dependent correction factors derived from the ratio of \textsc{HRes} (version 2.3) predictions at next-to-NLO (NNLO) accuracy obtained with and without finite quark mass effects included~\cite{deFlorian:2012mx}.
The \ggH signal is normalized to the cross section computed at next-to-NNLO in QCD, including NLO EW corrections~\cite{DAWSON1991283,DJOUADI1991440,PhysRevLett.70.1372,Spira:1995rr,Anastasiou:2002yz,Harlander:2002wh,Ravindran:2003um,Aglietti:2004nj,Actis:2008ug,Actis:2008ts,Anastasiou:2008tj,Harlander:2009bw,Harlander:2009mq,Pak:2009dg,Harlander:2009my,Anastasiou:2014vaa,Anastasiou:2014lda,Anastasiou:2015vya,LHCHiggsCrossSectionWorkingGroup:2016ypw,Karlberg:2024zxx}.
The VBF and \VH signals are normalized to cross sections computed at NNLO in QCD with NLO EW corrections~\cite{Ciccolini:2007jr,Ciccolini:2007ec,Denner:2014cla,Cacciari:2015jma,Brein:2003wg,Denner:2011id,Brein:2012ne,Altenkamp:2012sx,Harlander:2013mla,Harlander:2014wda,LHCHiggsCrossSectionWorkingGroup:2016ypw,Karlberg:2024zxx}.
All signal processes are also scaled to account for the \HTT branching fraction~\cite{Djouadi:1997yw,Spira:1997dg,Bredenstein:2006rh,Djouadi:2006bz,Bredenstein:2006ha,Denner:2011mq,LHCHiggsCrossSectionWorkingGroup:2016ypw}.

The description of the decay of the Higgs boson to \PGt leptons is obtained using the \PYTHIA generator (version 8.306)~\cite{Bierlich:2022pfr}. 
These samples are simulated without accounting for the \PGt lepton spin correlations. 
After the samples have been generated, the \TAUSPINNER package~\cite{Przedzinski:2018ett} is used to calculate event weights that can be applied to the simulated 
signal samples to model the expected \PGt lepton spin effects in the \HTT decay for \CP-mixing angles of $0^{\circ}$, $45^{\circ}$, and $90^{\circ}$.
Every other value of \alphacp can be modelled using a linear sum of these contributions. These weights do not affect the modelling of the \CP properties of the Higgs boson production interaction.
 
The \MGvATNLO~\cite{Alwall:2014hca} generator (version 2.9.13) is used to simulate the \Zjets and \Wjets backgrounds.
The \Zjets samples with matrix elements including up to two QCD-induced outgoing partons at the Born level are generated at NLO in QCD, and are merged using the FxFx scheme~\cite{fxfx}.
Additional \Zjets samples with electroweak production of two outgoing partons are generated at leading order (LO).
The \Wjets samples are generated at LO with matrix elements including up to four outgoing partons, and are combined using MLM jet matching~\cite{Alwall:2007fs}.
The \PYTHIA generator is used to simulate diboson production at LO. 
The \Zjets, \Wjets, and diboson processes are normalized to their theoretical cross sections, calculated at NNLO in QCD and including NLO EW corrections~\cite{Grazzini:2017mhc,Grazzini:2019jkl}.
The \POWHEG generator is used for top quark-antiquark pair production (\ttbar)~\cite{Alioli:2011as} and single top quark production (including the $t$-channel and $\mathrm{t}\PW$ associated production)~\cite{Re:2010bp, Frederix:2012dh} at NLO in QCD. 
The \ttbar and single-top $t$-channel processes are normalized to the cross sections computed at NNLO QCD~\cite{Czakon:2011xx, Campbell:2020fhf}, while the $\mathrm{t}\PW$ process is normalized to the cross section computed at NLO in QCD~\cite{Kidonakis:2021vob}.

All samples are generated with the NNPDF3.1~\cite{Ball_2017} NNLO parton distribution functions. 
The generators are interfaced with \PYTHIA to model the parton showering and fragmentation, as well as the decay of the \PGt leptons. 
The \PYTHIA parameters that affect the description of the underlying event are set to the {CP5} tune~\cite{Sirunyan_2020_CP5}.
Monte Carlo generated events are processed through a simulation of the CMS detector that is based on \GEANTfour~\cite{Agostinelli:2002hh}, and are reconstructed with the same algorithms as the ones used for data. The effect of pileup is taken into account by generating concurrent inelastic collision events with \PYTHIA. The pileup distribution in simulation is weighted to match that in data.

\section{Event selection}\label{sec:eventselection}

Events are selected online by the CMS trigger system. We consider \tauhtauh, \mutau, and \etau pairs for the measurement of \alphacp (\taultaul pairs are not included because of their low \CP sensitivity and small branching fraction). For the \ltau channels, events are triggered by a single-lepton trigger. For the \tauhtauh channel, a di-\PGt trigger and a di-\PGt plus jet trigger are used.

Offline, a pair of oppositely charged \PGt leptons separated by $\DR = \sqrt{\Deltaeta^2 + \Deltaphi^2} >0.5$ is required. The offline-reconstructed objects must match the required trigger objects (\ie the object as reconstructed by the trigger system) within $\DR<0.5$. The offline-reconstructed light lepton is required to have a \pt value that is at least 2\GeV higher than the online threshold. If an offline \tauh candidate is matched to a \tauh trigger object, the \tauh must have a \pt at least 5\GeV above the trigger threshold. 
The offline thresholds are higher than the online thresholds to exclude the \pt range in which the trigger efficiency is rising sharply.
Table~\ref{tab:triggertable} summarizes the online trigger and offline \pt thresholds used in the analysis. 

\begin{table}[htbp]
\renewcommand{\arraystretch}{1.5}
\centering
\topcaption{Kinematic trigger and offline requirements applied to the \etau, \mutau, and \tauhtauh channels. The \pt requirement is indicated in parentheses (in \GeVns). The \pt thresholds indicated for the jet apply only for the object matched to the jet leg of the ditau plus jet trigger.
\label{tab:triggertable}
}
\begin{tabular}{llll}
\hline

  Channel  & Trigger & Trigger requirement (\GeVns{})          &Offline requirement (\GeVns{})                       \\ 
  \hline 
 \multirow{2}{*}{\tauhtauh}  & ditau   &  $\tauh \,(35)\,\&\,\tauh \,(35)$             & $\tauh \,(40)\,\&\,\tauh \,(40)$        \\
              & $\text{ditau} + \text{jet}$ &  $\tauh \,(30), \tauh \,(30)\, \&\ \textrm{jet} \,(60) $  & $\tauh \,(35), \tauh \,(35)\, \&\ \textrm{jet} \,(60) $     \\
\mutau & single \PGm   & $\PGm \,(24)$\      & $\PGm \,(26)$    \\
\etau  & single \Pe  &  $\Pe \,(30)$\       & $\Pe \,(32)$   \\
\hline
\end{tabular}
\end{table}

For the \ltau channels, the large \Wjets background is reduced by rejecting events if the transverse mass \mT of the light lepton and \ptvecmiss system,
\begin{linenomath}
\begin{equation}
\mT \equiv \sqrt{{2 \pt^{\ell} \ptmiss [1-\cos(\Deltaphi)]}} < 65\GeV,
\label{eq:transversemass}
\end{equation}
\end{linenomath}
where \Deltaphi is the azimuthal angle between the direction of the light lepton and \ptvecmiss.

The longitudinal and transverse impact parameters \dz and \dxy of the muon and electron are required to satisfy $\abs{\dz}<0.2\cm$ and $\abs{\dxy}<0.045\cm$. 
For the leading \tauh track, only the requirement $\abs{\dz}<0.2\cm$ is imposed to avoid loss of selection efficiency. 
Further, a veto on events containing loosely identified additional electrons or muons is imposed to reduce the background from processes with additional leptons. 
In the small fraction of events where multiple \PGt lepton pairs are present, the pairs are ranked based on the output scores of the \textsc{DeepTau} algorithm for the \tauh candidates, and the relative isolation for the \taul candidates. 
The highest ranked pair is selected. 

The impact parameter significance (\SIP), defined as the magnitude of the impact parameter divided by its uncertainty, is required to satisfy $\SIP>1.0$ for the muon and electron candidates.
A cutoff of $\SIP>1.25$ is applied to the charged pion in \PGt decays via the $\pi$ decay mode. The selections on \SIP ensure that the impact parameter vector is well reconstructed.
Furthermore, the absolute magnitude of the relative energy difference between the charged pion and the neutral pion in the $\rho$ and \PaoneONEP decay modes is required to be greater than 0.2. This cutoff removes events for which the correct \phicp transformation required to account for the polarization of the intermediary meson is not clear.
These selections further increase the sensitivity of the analysis. 

\section{Background modelling}\label{sec:backgroundestimation}

The processes that contribute to the background in this analysis are \Zjets, \Wjets, \ttbar, single top quark, and diboson production.
Additionally, events comprised uniquely of jets produced through the strong interaction, referred to as QCD multijet events, form a significant background. 
These processes contribute to the production of genuine \PGt leptons, and to jets and leptons that are misidentified as \tauh, as well as prompt leptons and jets that are misidentified as \taul in the semileptonic channels. 
The majority of the backgrounds due to jets misidentified as \tauh candidates are estimated using the ``fake factor" (\FF) method, as described in Ref.~\cite{Sirunyan:2018qio}.
The remaining backgrounds, including the large genuine di-\tauh background, are determined using simulated events. 

\subsection{The \texorpdfstring{\FF}{fake factor} method}

This  method is designed to provide an estimate of the shape and normalization of events in which at least one quark or gluon jet is misidentified as a \tauh lepton based on control samples in data. We refer to such a jet as a \jettotauh. 

We define a determination region (DR) that is orthogonal to the signal region (SR) and dominated by a background process resulting in \jettotauh misidentifications.
We define a \tauh nominal identification (ID) as a \tauh object that passes nominal ID requirements as outlined in Section~\ref{sec:eventreconstruction}, and a relaxed \tauh ID as an object that passes a looser requirement on the \textsc{DeepTau} output but fails the nominal ID.
In this DR we obtain the ratio between the nominal ID \tauh rate and the relaxed ID \tauh rate. The ratio in the DR is the \FF.
To obtain the rate of misidentified jets in the SR, an application region (AR) is defined by selecting events that fulfil all event selection criteria except that they
 contain a \tauh lepton that passes the relaxed instead of the nominal requirement. 
The rate of misidentified jet events in the SR is obtained by applying the \FF values from the DR on an event-by-event basis as an event weight to the events in the AR.
In both DR and AR, the contribution of other background processes not involving \jettotauh events is subtracted using simulated events.
   
The \jettotauh background in the \tauhtauh channel originates almost entirely from QCD multijet events. 
The QCD multijet process produces \PGt lepton pairs with opposite-sign and same-sign charges at approximately the same rates (${\approx}45\%$ are same-sign), whereas the probability of producing same-sign events from genuine \PGt lepton pairs is small (${\approx}1.5\%$). 
The DR is thus defined by inverting the opposite-sign requirement on the \PGt lepton pair to a same-sign requirement.
The \FF are parameterized for the leading \tauh lepton as a function of the \pt of the \tauh, and binned in the reconstructed decay mode and jet multiplicity. 
The events in which the subleading \tauh is a misidentified jet and the leading \tauh candidate is a genuine \PGt lepton are modelled via simulation. These events constitute only a small fraction (${\approx}4\%$) of the total misidentified jet background in the \tauhtauh channel, and the simulation agrees with data within ${\approx}10\%$.

In the \ltau channels, the \Wjets process, and to a lesser extent the \ttbar process, contribute to jet misidentification, as well as events originating from QCD multijet production.
Therefore, separate \FF are derived for these processes, and these individual \FF values are subsequently weighted into an overall \FF to account for their different contributions in the AR.
Simulated events are used to determine the expected relative contributions of \Wjets and \ttbar events, and the QCD contribution is estimated by subtracting all simulated non-QCD processes from the data in the AR. 
The procedure for the QCD \FF is similar to the method described for the \tauhtauh channel, except that the light lepton relative isolation must be larger than 0.05 to reduce processes resulting in genuine leptons. 
A DR sufficiently pure in \Wjets is defined by selecting events with $\mT>70\GeV$ and using a BDT to reduce genuine \tauh backgrounds. 
Correction factors are derived using simulated \Wjets events to account for differences in the \FF due to these modified selection requirements. 
For the \ttbar process, it is difficult to define a sufficiently pure region in data, and thus the \FF values are estimated from a simulated \ttbar sample.
In this case, correction factors are applied to account for differences in the \FF in data and simulation, by comparing the \FF values measured in data and simulation for \Wjets events. 

\subsection{Estimation of other backgrounds}

The remaining misidentified background, which includes prompt leptons misidentified as \taul or \tauh, and \taul misidentified as \tauh, is estimated using simulated events.
Simulated events are also used to estimate the genuine \PGt lepton processes, which include the dominant \ZTT background, and smaller backgrounds from \ttbar, single top quark, and diboson production.
In order to model the background processes in data well, various corrections need to be applied to the simulated samples. 

Muons and electrons are corrected for mismodelling of their trigger, identification, and isolation efficiencies. 
The \tauh candidates are corrected for mismodelling of their trigger, reconstruction, and identification efficiencies. 
The electron energy scale is adjusted in data and simulation using fits to the dilepton invariant mass around the Z boson peak, and the resolution
of the simulated electrons is also adjusted to match the data.
The simulated \tauh energy scales are corrected per decay mode to match the corresponding scales in data~\cite{CMS:2025kgf}. 
The \ltotauh misidentification rates are corrected in simulation based on measurements of \Zll events, and the energy scales are corrected in simulation to match the scale in data~\cite{CMS:2025kgf}.

The impact parameters of charged particles are corrected using a quantile mapping technique. 
This method maps samples from a reference distribution (simulation) to match those of a target distribution (data), by applying a cumulative distribution function (CDF) transformation that matches their quantiles, following the inverse-CDF construction of Ref.~\cite{Read:1999kh}.
The muon and pion impact parameters are corrected using \Zmm events, and electron impact parameters are corrected using \Zee events.  
Each impact parameter vector component $\text{IP}_\text{i}$ ($\text{i} \in \{x,y,z\}$) is transformed as:
\begin{equation}
\text{IP}^{\text{corr}}_\text{i} = F^{-1}_{\text{data}}(F_{\text{sim}}(\text{IP}_\text{i})),
\end{equation}
where $\text{IP}^{\text{corr}}_\text{i}$ ($\text{IP}_\text{i}$) is the corrected (uncorrected) impact parameter component, $F_{\text{sim}}$ is the CDF of the impact parameter component in simulation and $F^{-1}_{\text{data}}$ is the inverse CDF of the impact parameter component in data.

Jet energy scale corrections are applied to both data and simulated events as described in Section~\ref{sec:eventreconstruction}.
Recoil corrections to the \ptvecmiss are measured in \Zmm events and are applied to reduce the mismodelling of the simulated \Zjets, \Wjets, and Higgs boson samples \cite{CMS-PAS-JME-12-002}. 
The corrections are applied to the simulated events as a function of the vectorial difference of the measured \ptmiss and total \pt of the neutrinos originating from the decay of the \PZ, \PW, or Higgs boson. 

In the \Zjets simulation, the dilepton invariant mass and \pt spectra are corrected to better match the data. Their distribution is determined in data and simulation in dimuon events, and the simulated events are corrected to match the spectra in data. 
A correction is also applied to the top quark \pt spectrum in the \ttbar sample, using the procedure detailed in Ref.~\cite{Khachatryan_2015}.

After applying all corrections, we obtain a satisfactory description of the observables that we use to categorize events.

\section{Event categorization}\label{sec:eventcategorisation}

In order to enhance the sensitivity of this analysis, we build BDT discriminants for each channel to separate signal from genuine \PGt and misidentified background events.

This event categorization is formulated as a multiclass problem, the output is a set of scores for each event (one per class) that, by construction, sum to unity. 
Each score can, therefore, be interpreted loosely as the probability that an event belongs to a given class.
Since both the \ltau and \tauhtauh channels are dominated by backgrounds containing contributions from genuine \PGt and \jettotauh production, the discriminant is trained to categorize events in three classes:
\begin{itemize}
\item The ``Higgs" category is trained to distinguish events from the \ggH, VBF, and \VH samples from background events, which are reweighted by their cross sections before merging them into one sample. Events in this category are used to determine \alphacp.
\item The ``Genuine" category includes all background processes involving two genuine \PGt leptons.
\item The ``Mis-ID" category includes all background processes in which minimally one hadronic jet is misidentified as a \tauh lepton. This category also contains \ltotauh misidentified events for all channels and prompt light leptons in the \ltau channels.
\end{itemize}

To avoid biases from the different number of events available for each class, event weights are applied so that the sum of weights is the same for each class, i.e. all classes contribute to the training with uniform prevalence. The backgrounds used for the training are taken from simulated samples, except for QCD events, which are obtained from data using same-sign \PGt lepton pair candidates.

A separate training was performed for each channel (\tauhtauh, \mutau and \etau), using a multiclass BDT algorithm combined with the \textsc{XGBoost} package, with hyperparameters optimized using the \textsc{Optuna} framework~\cite{xgboost,optuna}. The training is performed inclusively for all the possible \PGt lepton decay modes within the given channel.  

The input variables used in the categorization of the \ltau and \tauhtauh channels include the \pt of the \tauh or lepton, the \ptmiss, and the \ptvecmiss azimuthal separation from each object in the pair. Further information about the pair is provided by including the visible di-\PGt mass and \pt, the di-\PGt mass and \pt as estimated by the \FASTMTT algorithm, the angular separation of the pair, and the \mT between each candidate and \ptvecmiss.
Additional variables related to jets in the event are used: the leading and subleading jet \pt, the dijet information (invariant mass, \pt and \Deltaeta), and the jet multiplicity. None of these variables are sensitive to the \CP nature of the \PGt coupling, and the BDT output distribution does not depend on the value of \alphacp.

We note that defining separate classes for different production mechanisms (\eg VBF+\VH and \ggH) does not improve the sensitivity to \alphacp, as the BDT already exploits the mechanism-specific variables to separate signal from background, pushing events from production modes with more distinctive kinematic signatures relative to the background, such as VBF, toward more signal-rich regions of the BDT output. 

Events are sorted into the three categories depending on which of the three output scores is largest. The maximum output score is also retained and used for the purpose of signal extraction. These maximum scores will be referred to as the ``BDT scores" in the following.   

\begin{figure}[!htb]
    \centering
       \includegraphics[width=0.49\textwidth]{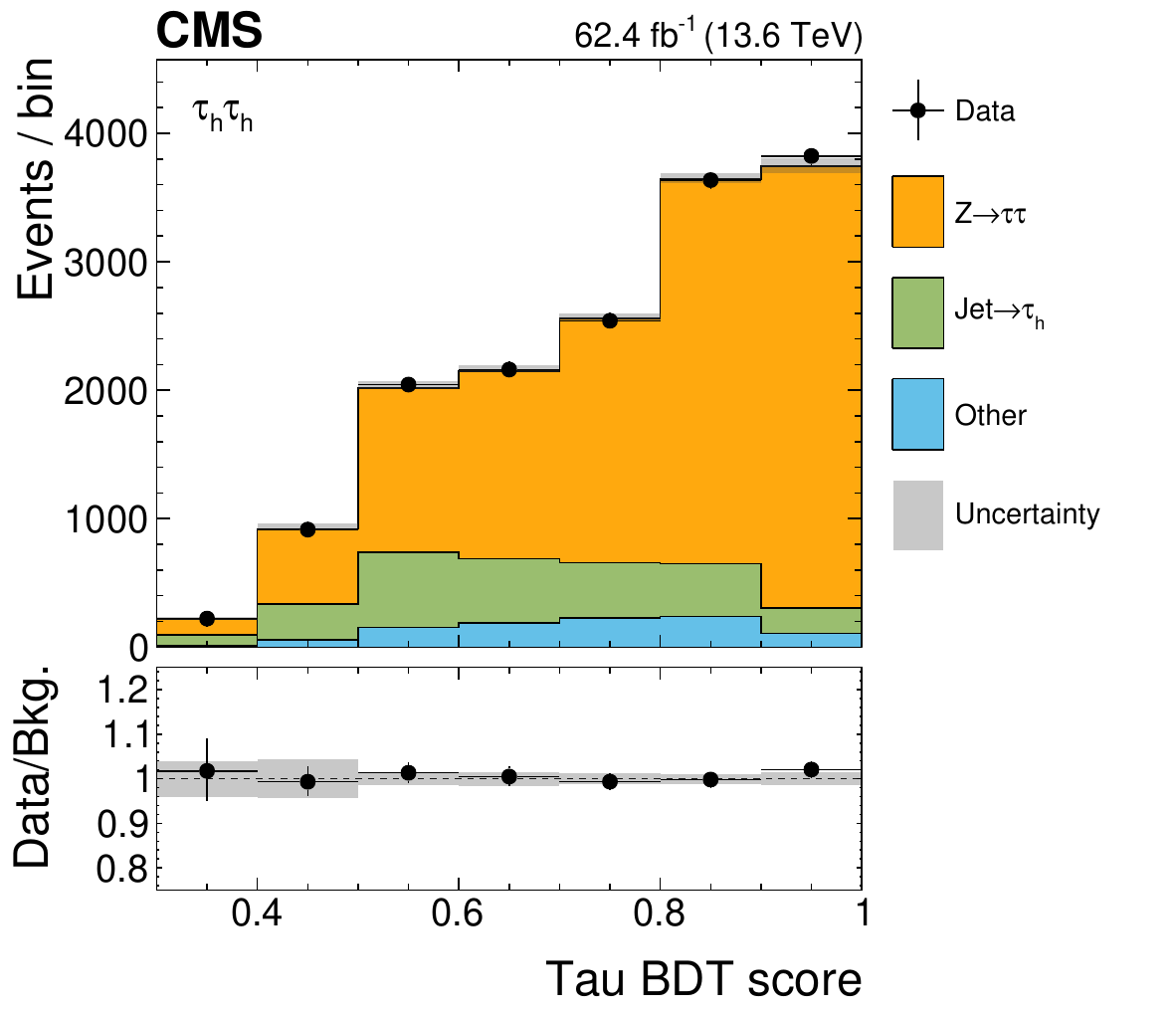}
       \includegraphics[width=0.49\textwidth]{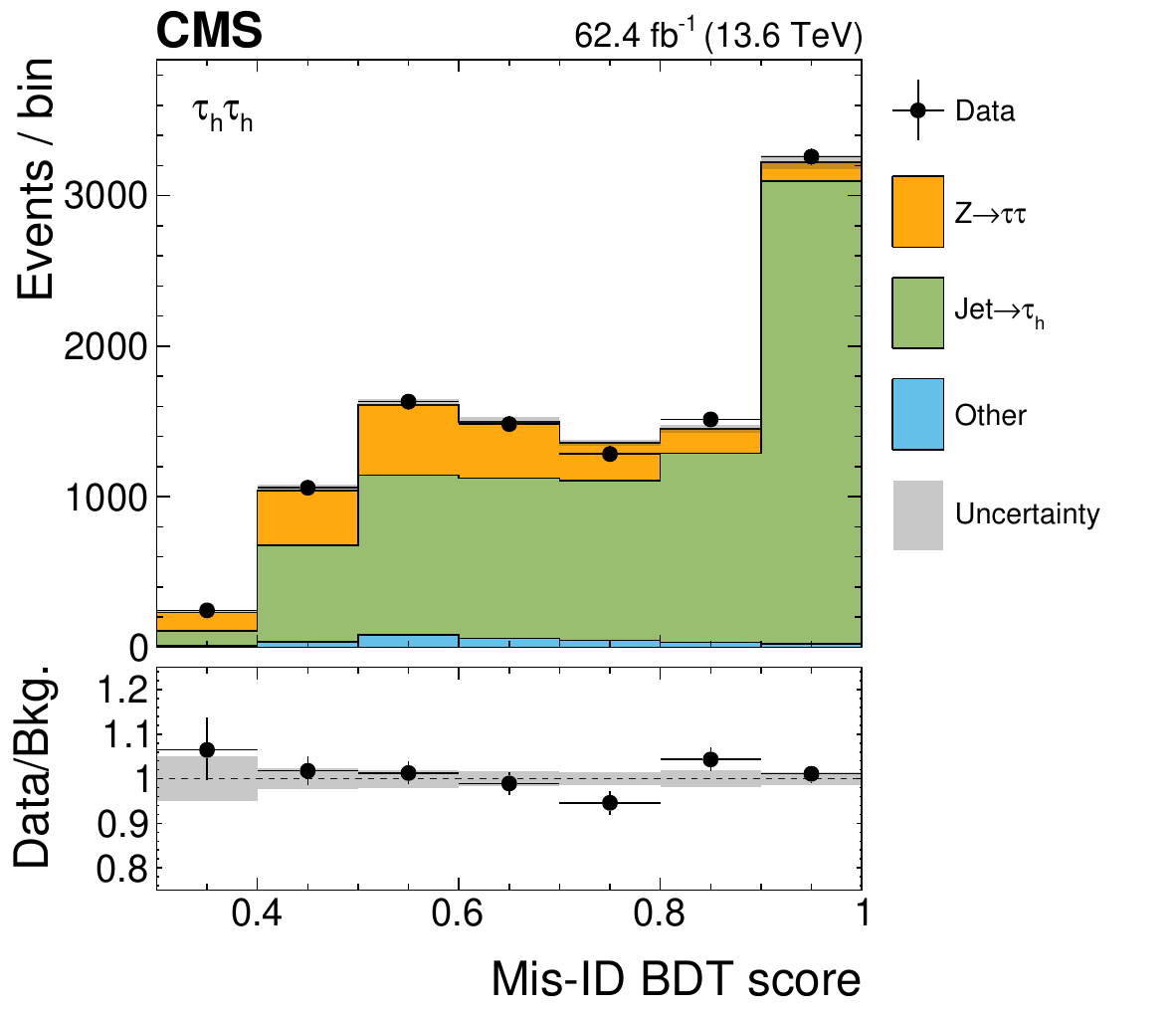}
    \caption{The post-fit BDT score distributions for the Genuine (left) and Mis-ID (right) categories in the \tauhtauh channel. The distributions are inclusive in \tauh decay mode.  In the lower panels, the data divided by the expectation is displayed. The uncertainty band accounts for all sources of systematic uncertainty in the background prediction. \label{fig:tt_BDT}}
\end{figure}

The BDT score distributions, after the fit to data described in Section 12, of the Genuine and Mis-ID categories are displayed for the \tauhtauh channel in Fig.~\ref{fig:tt_BDT}. 
The genuine di-\PGt and \jettotauh background contributions are displayed separately as indicated in the legends. 
The remaining background contributions are collated and indicated by the ``Other" label.

\section{Acoplanarity angle distributions }\label{sec:unrolled}

The effective mixing angle \alphacp is extracted by fitting the \phicp distributions.
The BDT score distributions described in Section~\ref{sec:eventcategorisation} allow for a partial separation of signal from background events. 
The events assigned to the Higgs category by the BDT are then separated by the final state of the \PGt lepton pair, defining several signal categories.
The \phicp distributions of the events in the signal categories are then analysed in windows of increasing Higgs category BDT score, corresponding to progressively higher signal-to-background ratios.
The result is a set of two-dimensional distributions built from the BDT score and \phicp variables, which are used in the fit to data to extract the results. 

The modelling of the \phicp spectrum was validated using the procedure described in Section 8.4 of Ref.~\cite{run2_cp_paper} (originally proposed in Ref.~\cite{Berge:2014sra}). This involves dividing the \mutau events into distributions in which the charged pion is nearly parallel or nearly perpendicular to the production plane. This yields two \phicp distributions with modulations of opposite phase, which are found to be well modelled.

The statistical fluctuations in the estimates of the background contributions (denoted as background templates) in the signal and background categories are sizeable.
The \phicp distribution for backgrounds involving two genuine \PGt leptons, as well as the \jettotauh background are symmetric around $\phicp=\pi$. These backgrounds are therefore symmetrized, meaning the symmetry around $\phicp=\pi$ is enforced to reduce the impact of the statistical fluctuations. 

The signal templates for the scalar and pseudoscalar cases are symmetrized around $\phicp=\pi$ as well. 
The maximally mixed signal template, which is not displayed in the plots, is used in the fitting procedure described in Section~\ref{sec:results}. 
In order to reduce the impact of the statistical fluctuations in this template, we reweight the signal sample to another sample with $\alphacp=-45^{\circ}$. 
The \phicp distribution is shifted by $\pi$, and the average is taken between the samples with $\alphacp=45^{\circ}$ and $\alphacp=-45^{\circ}$. 

\begin{figure}[!htb]
    \centering
     \includegraphics[width=1\textwidth]{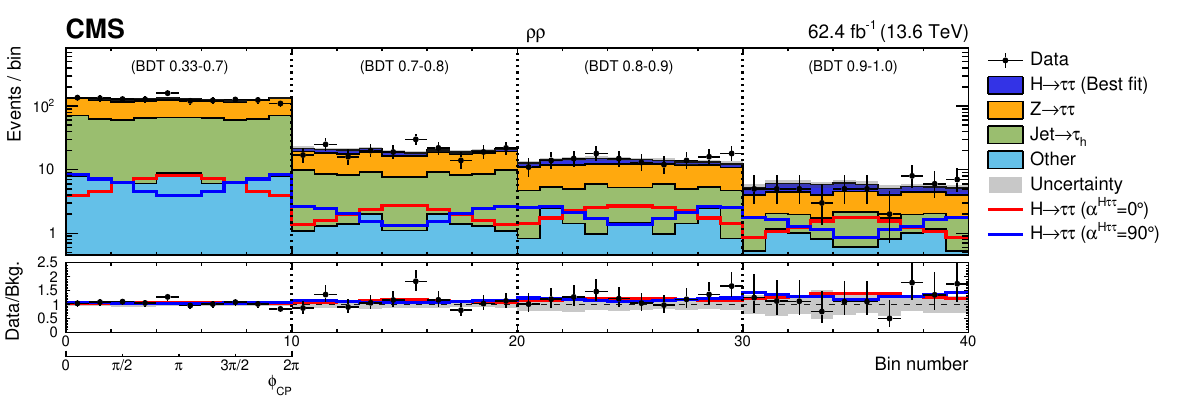}
     \includegraphics[width=1\textwidth]{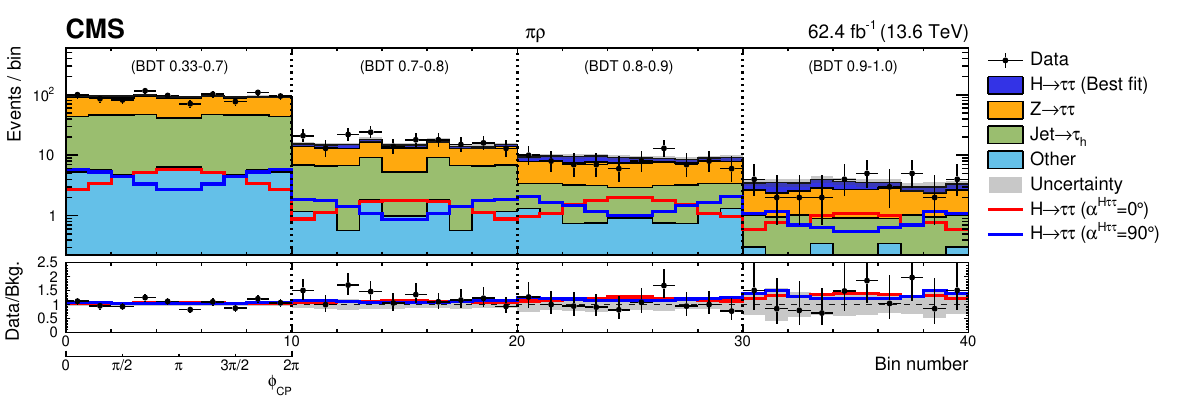}
     \includegraphics[width=1\textwidth]{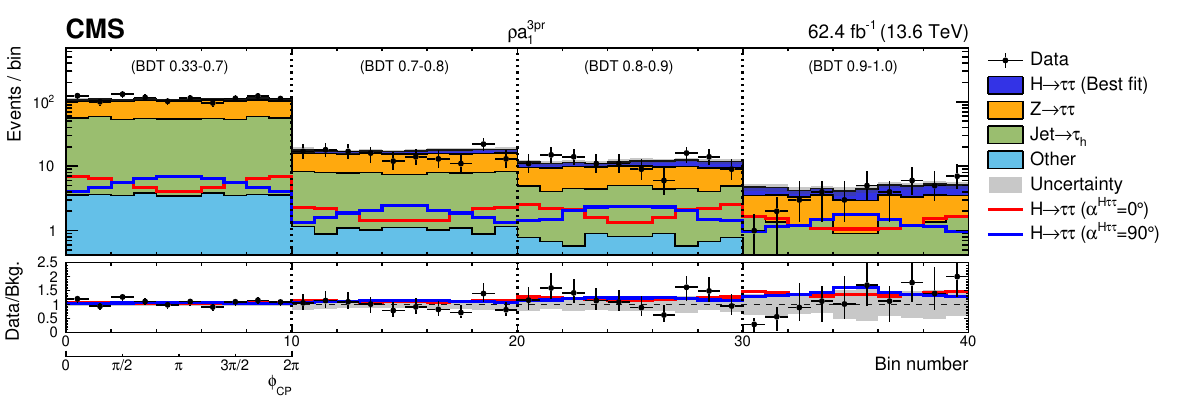}
\caption{From upper to lower: distributions of \phicp in the \rhorho, \pirho, and \rhoaone channels in windows of increasing Higgs category BDT score. The scalar ($\alphacp=0^\circ$) and pseudoscalar ($\alphacp=90^\circ$) signal scenarios are overlaid, with signal normalization taken from the fit to data presented in Section \ref{sec:results}.
In the lower panels, the data divided by the background expectation is displayed, as well as the scalar and pseudoscalar signal-plus-background expectations divided by the background. The uncertainty band accounts for all sources of systematic uncertainty in the prediction. }
\label{fig:unrolled}
\end{figure}

Figure ~\ref{fig:unrolled} shows the data, signal and background distributions, with the scalar and pseudoscalar signal templates overlaid, for the three most sensitive decay mode combinations (\rhorho, \pirho, and \rhoaone). 
The distributions highlight the effectiveness of the BDT discriminant in optimizing the signal over background ratio, as well as the \CP-sensitivity of the measurement that follows from the visibly different phases of the \CP-even and \CP-odd signal distributions.

\section{Systematic uncertainties}\label{sec:systematicuncertainties}

The sensitivity of the measurement of \alphacp is dominated by statistical uncertainties. 
Systematic effects have a comparatively small impact, with a total uncertainty in \alphacp of approximately 5\% of the statistical uncertainty. 
Systematic uncertainties affecting the shapes and yields of signal and background processes are incorporated into the likelihood as nuisance parameters.

The largest contribution to the systematic uncertainty is due to the limited number of events in the signal and background templates.
In the signal categories, this is accounted for using the Barlow--Beeston method~\cite{Conway:2011in}, which assigns a single nuisance parameter per pair of symmetrized bins per process. 
For the background categories, the Barlow--Beeston ``lite" method~~\cite{BarlowBeeston,Conway:2011in} is used, which assigns a single nuisance parameter per bin that scales the total background yield in that bin.

Theoretical uncertainties in the signal yield originate from missing higher-order QCD contributions, estimated by varying the QCD renormalization and factorization scales in the MC generators, and uncertainties in the parton shower modelling. 
Theoretical uncertainties in the modelling of the \ttbar background originate from the uncertainty in the top quark \pt spectrum.
Uncertainties in cross sections and branching fractions of the signal and background processes are also considered but have a minor impact on the final result.

Experimental uncertainties arise from the reconstruction and identification efficiencies and energy scale calibrations of the physics objects, primarily the \tauh leptons, with subleading effects due to the muons and electrons. 
Uncertainties related to the modelling of the \Zjets di-lepton invariant mass are evaluated by comparing the simulation to data in a \Zmm control region.
Other leading systematic uncertainties include the jet and \ptmiss scale and resolution uncertainties. 

We also consider uncertainties in the $\vec{\textrm{P}}^\pm$ and $\vec{\textrm{R}}^\pm$ vectors that affect the \phicp distributions directly.
The largest of these uncertainties originates from the modelling of the impact parameters and SV resolutions, which we estimate by comparing data and simulation as a function of angular variables that are sensitive to these effects.
The overall impact on \alphacp is small.
The uncertainties in other quantities used in the \phicp calculation were found to have a negligible impact on the final result and were therefore not included.

Uncertainties in the \FF method used to estimate the \jettotauh background also have a small contribution to the total uncertainty in \alphacp. 
These uncertainties include statistical uncertainties in the \FF values, as well as systematic uncertainties which are evaluated using closure tests using simulated events and/or data control regions.
The uncertainties in \ltotauh misidentification rates and energy scales are also included, but have a minor impact on the final result as the corresponding backgrounds are small.
Other uncertainties, including those in the integrated luminosity and the pileup reweighting, have a negligible impact on the final result.

\section{Results}\label{sec:results}

In order to extract the \CP-mixing angle \alphacp, a fit to the data is performed using the simultaneous binned likelihood function \likeli, that is a function of the \CP-mixing angle \alphacp, the Higgs boson signal strength modifiers \muggh and \muqqh (defined as the production cross section times the \HTT branching fraction divided by the corresponding SM prediction), and the nuisance parameters \vectheta that account for the systematic uncertainties. In the fit, all \HTT production processes involving \PV boson couplings, namely VBF and \VH, are scaled by \muqqh, while the \ggH process is scaled by \muggh. 
The use of two signal strength modifiers in the fit reduces the dependence of the result on assumptions about the relative contributions of different production processes.
The following results have been determined using the CMS statistical analysis tool \textsc{Combine}~\cite{CMS:2024onh}.

The likelihood function is defined as a product of conditional probabilities $P$ over binned distributions of the discriminating observables in each event category:
\begin{linenomath}
\begin{equation}
    \likeli = 
    \prod^{N_{\text{bin}}}_{i} P(\nij \mid \Si + \Bi) \, \prod^{N_{\text{nuisance}}}_{m} \Cm.
  \label{eq:likelihood}
\end{equation}
\end{linenomath}
In this equation the index i ranges over all bins in all analysis categories. The Poisson distributions $P$ correspond to the observation of \nij events in each bin of the discriminating observable given the expectation for the background \Bi and the signal \Si.
Constraints on the nuisance parameters corresponding to the systematic uncertainties are represented by the functions \Cm. A more detailed discussion on the formulation of the statistical inference may be found in Refs.~\cite{CMS-NOTE-2011-005,Conway:2011in}.
The systematic uncertainties affecting the normalization of the signal and background templates are incorporated in the fit via nuisance parameters with a log-normal constraint. The shape-altering systematic uncertainties are represented by nuisance parameters whose variations cause continuous morphing of the signal or background template shapes, and are assigned a Gaussian constraint. The bin-by-bin statistical uncertainties in the background samples are also assigned a Gaussian constraint.

The inputs to the likelihood fits differ for the signal and background categories. For the signal categories, the \phicp distributions in bins of the BDT score are used. 
For the background categories, the BDT score distributions are used, which allows systematic uncertainties to be further constrained, and helps to improve the fit convergence.

\subsection{Results at \texorpdfstring{$\sqrt{s} = 13.6\TeV$}{sqrt(s) = 13.6 TeV}}

\begin{figure}[!htb]
    \centering
    \includegraphics[width=0.7\textwidth]{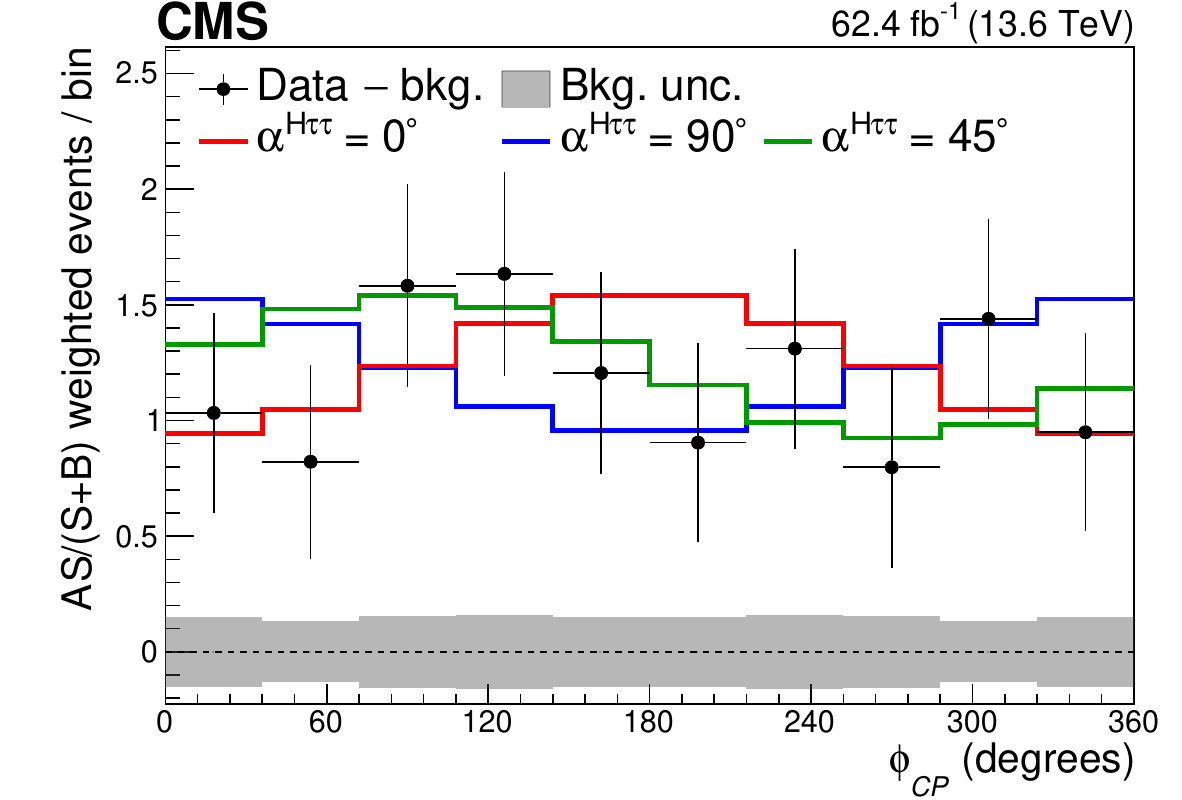}
    \caption{
    The \phicp distributions for the \rhorho, \pirho, \rhoaone, $\PGr\PaoneONEP+\PaoneONEP\PaoneONEP$, $\Pgm\PaoneTHREEP$, $\Pe\PaoneTHREEP$, \murho, and \erho channels in 13.6\TeV data, are weighted by \AveAsym and combined. 
    Events are included from all BDT score bins in the signal categories. 
    The background is subtracted from the data. 
    The \CP-even distribution is depicted in red, the \CP-odd in blue, and a mixed-\CP distribution ($\alphacp=45^{\circ}$) in green. 
    In the predictions, the cross section times branching fraction are taken from their best fit values. 
    The grey uncertainty band indicates the uncertainty in the subtracted background component. 
    In combining the channels, a phase shift was applied to the channels where the \phicp has a different phase with respect to the \rhorho channel.
    \label{fig:weighted_phicp_run3}}
\end{figure}

A fit is performed to the data collected at $\sqrt{s} = 13.6\TeV$ (62.4\fbinv).
Figure~\ref{fig:weighted_phicp_run3} shows the \phicp distribution for the \rhorho, \pirho, \rhoaone, $\PGr\PaoneONEP+\PaoneONEP\PaoneONEP$, $\Pgm\PaoneTHREEP$, $\Pe\PaoneTHREEP$, \murho, and \erho channels with the fitted background subtracted from the data, 
together with the \CP-even, \CP-odd, and mixed-\CP ($\alphacp=45^{\circ}$) predictions.
These channels are chosen as the same number of \phicp bins are used in the fit to data, and collectively they account for most of the sensitivity to \alphacp.
Events are included from all BDT score bins in these signal categories. Each BDT score bin is weighted by \AveAsym, where $S$ and $B$ are the signal and background rates, respectively, and $A$ is a measure for the average asymmetry between the scalar and pseudoscalar distributions. 
The definition of the value of $A$ for each BDT bin is:
\begin{equation}
    A = \frac{1}{N_{\phicp}} \sum^{N_{\phicp}}_{i} \frac{\abs{N_i^{\text{even}}-N_i^{\text{odd}}}}{N_i^{\text{even}}+N_i^{\text{odd}}},
\end{equation}
where $N_i^{\text{even}}$ and $N_i^{\text{odd}}$ are the number of events in each \phicp bin for the \CP-even and \CP-odd signal respectively, and $N_{\phicp}$ is the total number of \phicp bins within the BDT bin. 
Compared with the predictions, the measured \phicp distribution has a somewhat smaller modulation amplitude, and shows a slight preference for the mixed-\CP hypothesis relative to the pure \CP-even and \CP-odd cases. 
The reduced amplitude decreases the sensitivity to \alphacp, leading to a larger observed uncertainty, as discussed below. 
The preference for the mixed-\CP hypothesis is not statistically significant, and the data remain compatible with the \CP-even scenario at the 1.2~$\sigma$ level.

\begin{figure}[!htb]
    \centering
    \includegraphics[width=0.7\textwidth]{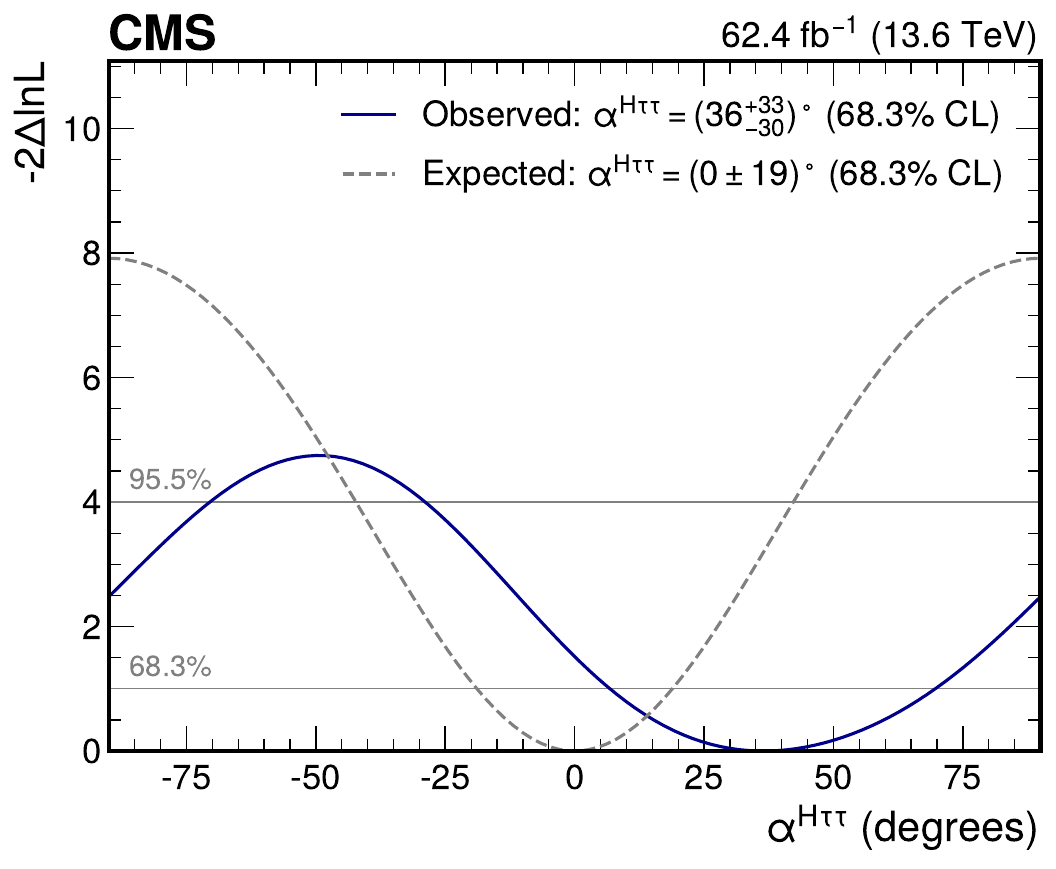}
    \caption{
        Negative log-likelihood scan for the combination of the \tauhtauh, \mutau, and \etau channels in 13.6\TeV data. The best fit value of \alphacp is found to be $(36^{+33}_{-30})^\circ$, compared with an expected value of $(0 \pm 19)^\circ$.
    \label{fig:LLscan}}
\end{figure}

A negative log-likelihood scan for the combination of the \etau, \mutau, and \tauhtauh channels is shown in Fig.~\ref{fig:LLscan}. 
The value of \alphacp is found to be $(36^{+33}_{-30})^\circ$, compared with an expected value of $(0 \pm 19)^\circ$, both at the 68.3\% confidence level (\CL).
The observed uncertainty is larger than expected because the reduced modulation amplitude of the measured \phicp distribution, discussed above, lowers the sensitivity to \alphacp. 
The probability of obtaining an uncertainty larger than the one measured was estimated using pseudoexperiments to be approximately 6\%, and we thus conclude that the observed uncertainty is compatible with a statistical fluctuation.
The parameters that scale the ggH and qqH production signal strength with respect to the SM predictions (\muggh and \muqqh) are left to float freely in the fit. The best fit values of these parameters are $\muggh = 1.86^{+0.48}_{-0.47}$ and $\muqqh = -1.3^{+1.1}_{-1.4}$ respectively. 
These two parameters are strongly anticorrelated, with a correlation coefficient of $-0.76$, as the analysis does not directly attempt to differentiate between production modes.
The fit prefers a negative value of \muqqh, this is largely due to the strong anticorrelation between the two signal strength parameters, and therefore compensated by \muggh being larger than the expected value.
The compatibility of the data with the SM was determined, from pseudoexperiments, to have a $p$-value of 13\%.
The total cross section, measured to be $0.94^{+0.28}_{-0.22}$ times the SM expectation, is compatible with the SM. 
An alternative fit using a single signal strength parameter for all production modes yields a compatible result for \alphacp.

The observed uncertainty in \alphacp can be decomposed into statistical, bin-by-bin fluctuations in the background and signal templates, and experimental and theoretical systematic uncertainties. In this decomposition we obtain:
\begin{equation*}
    \alphacp = (36 \pm 27 \, (\text{stat}) \, {}^{+18}_{-11} \, (\text{bin-by-bin}) \, {}^{+9}_{-7} \, (\text{syst}) )^\circ.
\end{equation*}
The largest component of the systematic uncertainty originates from the bin-by-bin fluctuations in the templates.

The expected decomposed uncertainty is $\alphacp = (0 \pm 19 \, (\text{stat}) \, \pm 2 \, (\text{bin-by-bin}) \, \pm 2 \, (\text{syst}))^\circ$. The larger observed values of the bin-by-bin, experimental and theoretical systematic uncertainties are partially due to the reduced modulation amplitude observed in data, discussed above, and a consequence of the non-zero best fit value. The latter is due to the nuisance parameters having mostly symmetric effects on the templates (about $\phicp=\pi$), and therefore the impact on the uncertainty is smaller when the signal is symmetrical (i.e. closer to \CP-even or \CP-odd) than when it is asymmetrical (i.e. closer to \CP-mixed).

The expected exclusion of the pure \CP-odd scenario assuming the SM hypothesis ($\alphacp=0^\circ$) is $2.8\sigma$, showcasing the increased sensitivity of this analysis with respect to the 13\TeV result~\cite{run2_cp_paper}, which uses more than twice the amount of data (138\fbinv) but achieves a lower expected sensitivity of $2.6\sigma$. The expected sensitivity for the 13\TeV measurement using 62.4\fbinv of data would be $1.7\sigma$.
This improvement can be attributed to several factors. The signal acceptance was increased by approximately 40\% due to improved tau lepton identification and trigger algorithms. Additionally, the Higgs boson production cross section increased by approximately 10\% at the higher $\sqrt{s} = 13.6$~\TeV relative to $\sqrt{s} = 13$~\TeV. The sensitivity of the measurement was further improved by approximately 20\% through optimization of the analysis selections described in Section~\ref{sec:eventselection}. This included the introduction of a cutoff on the relative energy difference between pions, an adjusted \SIP selection, a new threshold for tau identification, and improved decay mode assignment. Furthermore, the use of the polarimetric vector for final states involving a single $\PaoneTHREEP$ decay was introduced for the first time in this work, increasing analysis sensitivity by a further 10\%.

The most sensitive channel is \tauhtauh with an expected exclusion of the pure \CP-odd scenario of $2.4\sigma$, followed by the \mutau and \etau channels with expected exclusions of $1.3\sigma$ and $0.7\sigma$, respectively.

\subsection{Combination of \texorpdfstring{$\sqrt{s} = 13$ and $13.6\TeV$}{sqrt(s) = 13 and 13.6 TeV} data sets}

This analysis is combined with the measurement performed at $\sqrt{s} = 13\TeV$ from Ref.~\cite{run2_cp_paper}, which observed a value of $\alphacp = (-1 \pm 19)^\circ$ compared with an expected value of $(0 \pm 21)^\circ$.

\begin{figure}[!htb]
    \centering
    \includegraphics[width=0.7\textwidth]{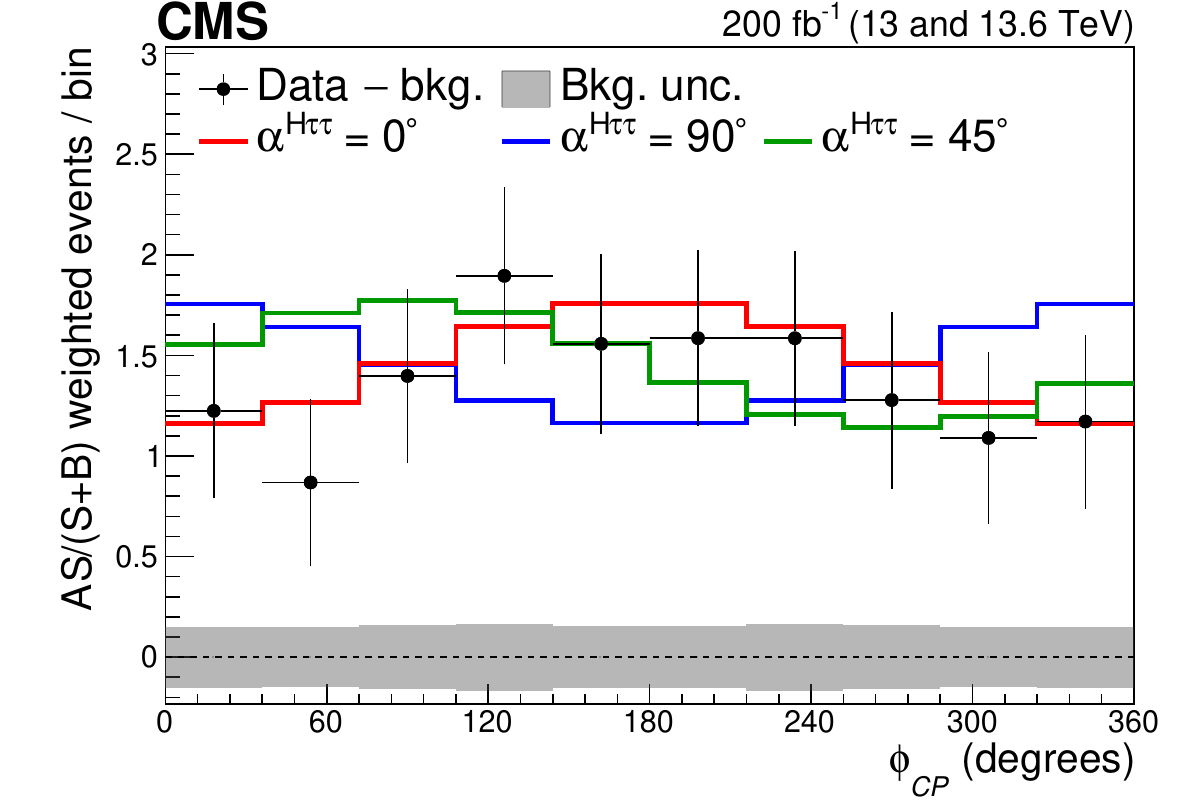}
    \caption{
    The \phicp distributions for the \rhorho, \pirho, \murho, and \erho channels in 13 and 13.6\TeV data, and the \rhoaone, $\PGr\PaoneONEP+\PaoneONEP\PaoneONEP$, $\Pgm\PaoneTHREEP$, and $\Pe\PaoneTHREEP$ channels in 13.6\TeV data, are weighted by \AveAsym and combined. 
    Events are included from all BDT score bins in the signal categories. 
    The background is subtracted from the data. 
    The \CP-even distribution is depicted in red, the \CP-odd in blue, and a mixed-\CP distribution ($\alphacp=45^{\circ}$) in green. 
    In the predictions, the cross section times branching fraction are taken from their best fit values. 
    The grey uncertainty band indicates the uncertainty in the subtracted background component. 
    In combining the channels, a phase shift was applied to the channels where the \phicp has a different phase with respect to the \rhorho channel.
    \label{fig:weighted_phicp_cmb}}
\end{figure}

Figure~\ref{fig:weighted_phicp_cmb} shows the \phicp distribution for the \rhorho, \pirho, \murho, and \erho channels in the $\sqrt{s} = 13$~\TeV data, 
added to the $\sqrt{s} = 13.6$~\TeV data shown previously in Fig.~\ref{fig:weighted_phicp_run3}.
Compared with Fig.~\ref{fig:weighted_phicp_run3}, the combined data set shows a distribution that is more compatible with the \CP-even prediction.

\begin{figure}[!htb]
    \centering
    \includegraphics[width=0.7\textwidth]{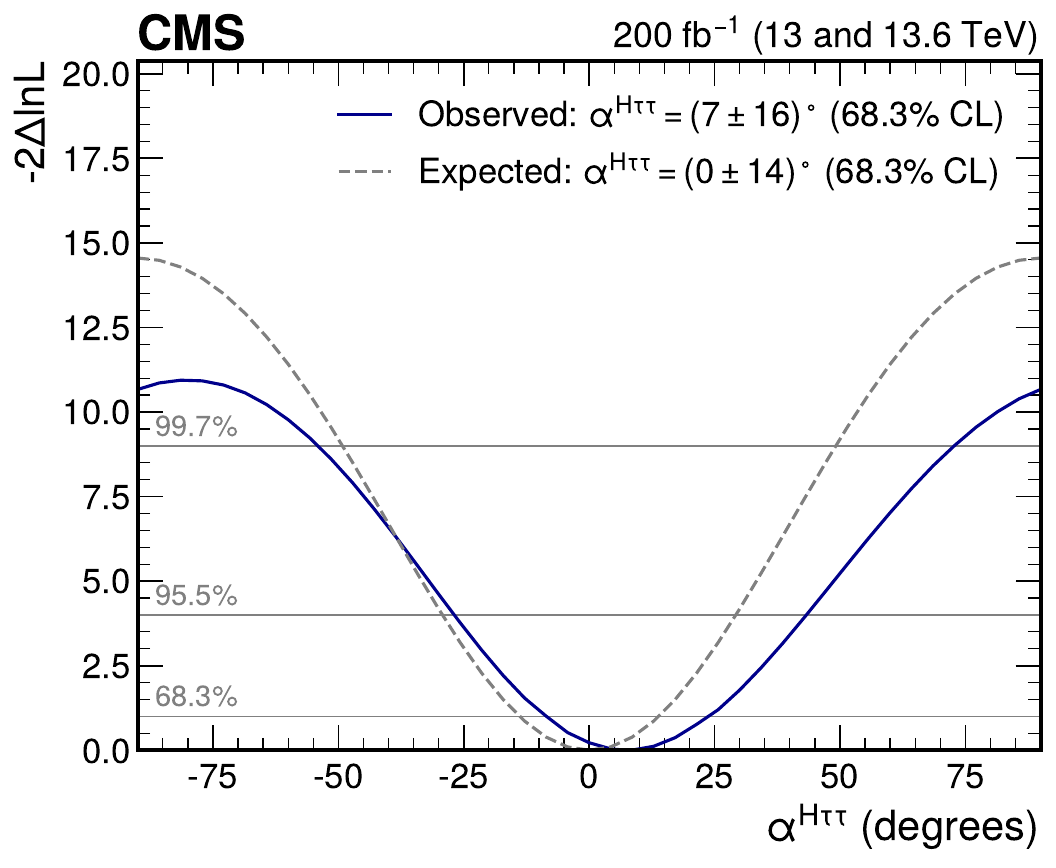}
    \caption{
        Negative log-likelihood scan for the combination of the \tauhtauh, \mutau, and \etau channels in 13 and 13.6\TeV data. The best fit value of $\alphacp$ is $(7 \pm 16)^\circ$ compared with an expected value of $(0 \pm 14)^\circ$.
    \label{fig:LLscan_combined}}
\end{figure}

The combined negative log-likelihood scan is shown in Fig.~\ref{fig:LLscan_combined}. 
The value of $\alphacp$ is determined to be $(7 \pm 16)^\circ$ compared with an expected value of $(0 \pm 14)^\circ$.
This represents the most precise measurement by CMS of the \CP nature of the Higgs boson coupling to \PGt leptons, with an expected precision that is the best achieved by any experiment to date.

\begin{figure}[htbp]
    \centering
    \includegraphics[width=0.5\textwidth]{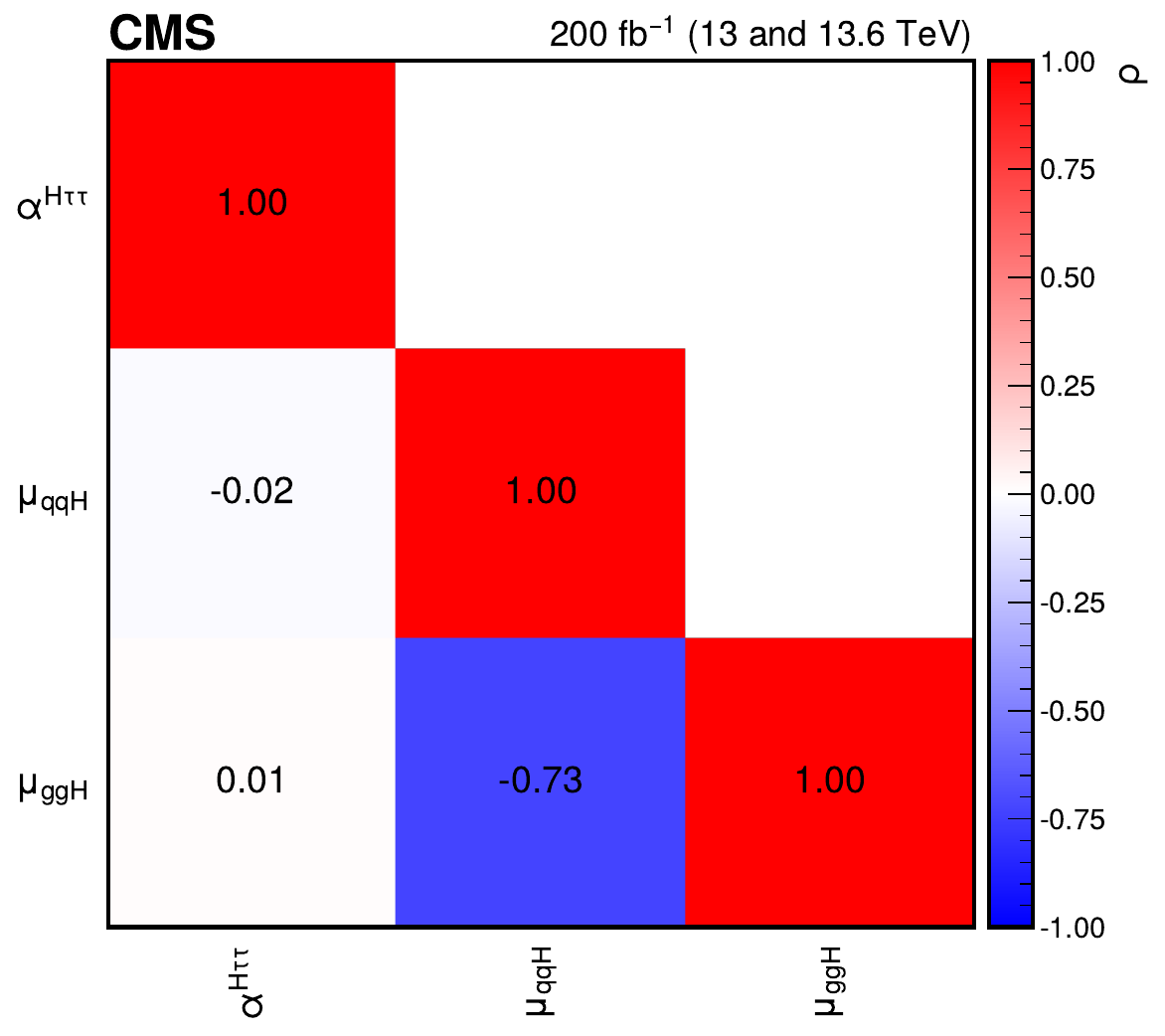}
    \caption{Correlation matrix for the mixing angle \alphacp and the \muggh and \muqqh signal strength modifiers, for the combined fit to the 13 and 13.6\TeV data sets.}
    \label{fig:corr_matrix}
\end{figure}

The best fit values of the signal strength parameters are $\muggh = 0.93^{+0.29}_{-0.25}$ and $\muqqh = 0.79^{+0.50}_{-0.47}$ respectively. 
These parameters remain strongly anticorrelated, with a correlation coefficient of $-0.73$, as shown by the correlation matrix in Fig.~\ref{fig:corr_matrix}. 
The \CP-sensitive parameter \alphacp is found to be essentially uncorrelated with the production signal strengths. 
Together with the best fit values and their uncertainties, the correlation matrix characterises the profiled likelihood in the Gaussian approximation, allowing constraints in the (\muggh, \muqqh) plane, or reinterpretations of the result within specific models, to be derived without the full multidimensional likelihood scan.

The observed uncertainty can be decomposed into:
\begin{equation*}
    \alphacp = ( 7 \pm 16 \, (\text{stat}) \, {}^{+3}_{-2} \, (\text{bin-by-bin}) \, {}\pm 2 \, (\text{syst}) )^\circ.
\end{equation*}
The total uncertainty is dominated by the statistical component. 

\begin{figure}[!htbp]
\centering
\includegraphics[width=0.6\textwidth]{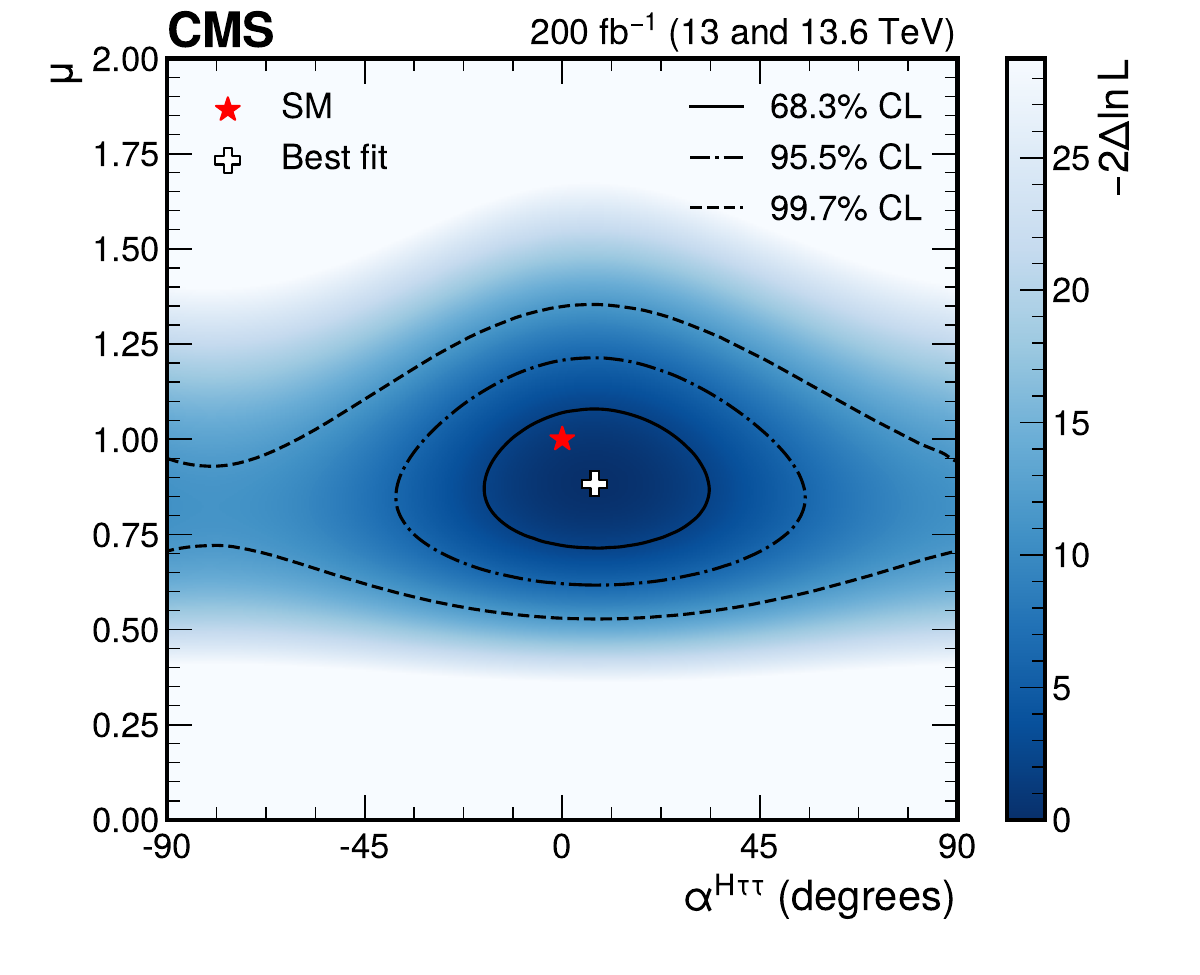}
\caption{Observed scan of $-2\Delta\ln L$ for $\mu$ against \alphacp for the combination of the 13 and 13.6\TeV measurements. The red marker indicates the SM prediction, and the white marker indicates the best fit to the data. }
\label{fig:muvsalpha}
\end{figure}

A fit to the data is also performed assuming \muggh = \muqqh = $\mu$. In this case $\mu$ is the combined signal strength modifier that scales the total H production cross section times $H\to\tau\tau$ branching fraction relative to the SM value. The best fit is for $\mu = 0.88^{+0.13}_{-0.12}$ and $\alphacp= (7 \pm 16 )^\circ$. A negative log-likelihood scan of $\mu$ versus \alphacp is shown on Fig.~\ref{fig:muvsalpha}.

\begin{figure}[htbp]
\centering
\includegraphics[width=0.6\textwidth]{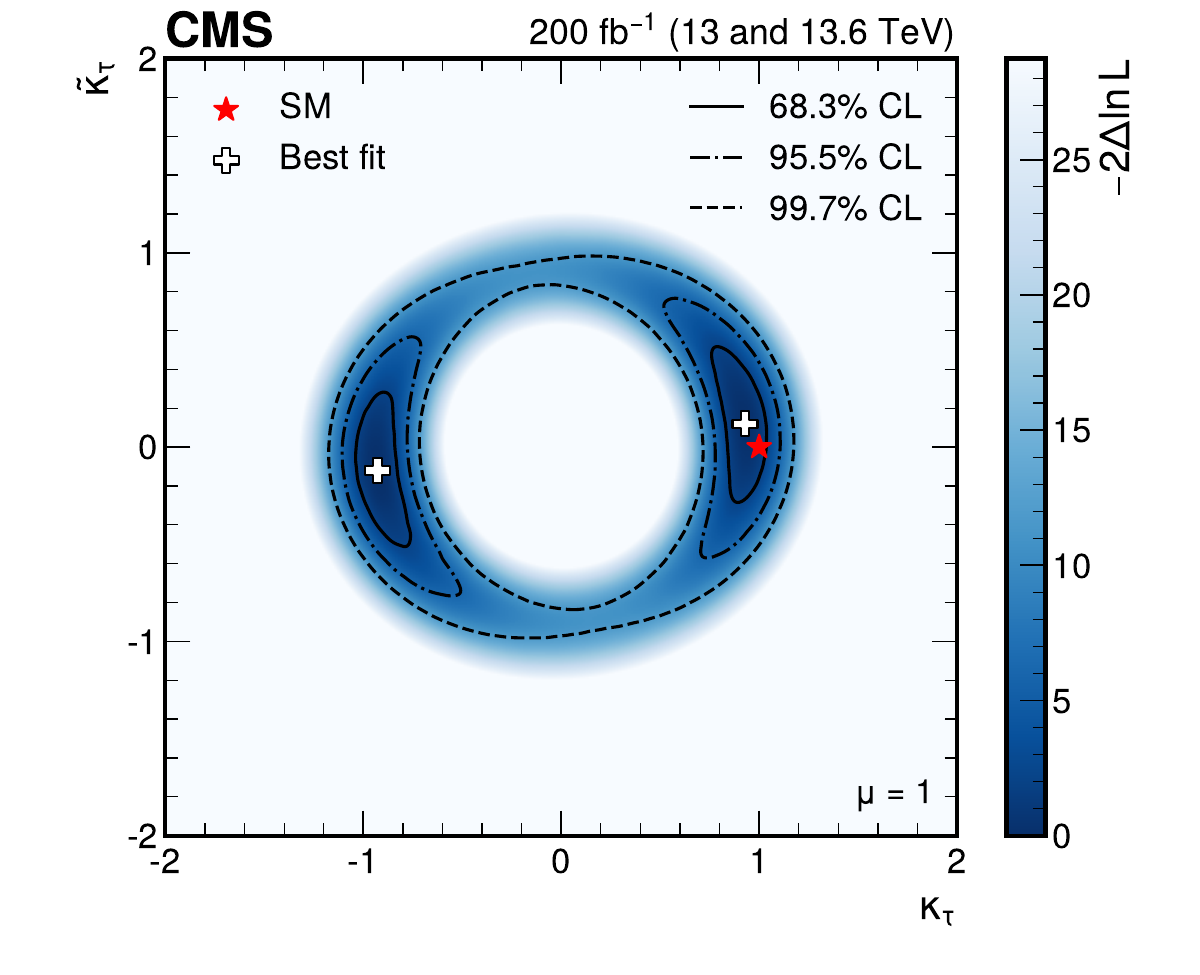}
\caption{Observed scan of $-2\Delta\ln L$ for the reduced \CP-odd (\kappattilde) coupling against the reduced \CP-even (\kappat) coupling for the combination of the 13 and 13.6\TeV measurements. The red marker indicates the SM prediction, and the white markers indicate the best fits to the data. }
\label{fig:kappas}
\end{figure}

A fit is also performed parameterizing the likelihood in terms of the \CP-even (\kappat) and \CP-odd (\kappattilde) couplings to \PGt leptons, where all other Higgs couplings are fixed to their expected SM values. 
This is shown in Fig.~\ref{fig:kappas}. The fit is only sensitive to the relative sign between \kappat and \kappattilde, and there are, therefore, two best fit points.

\subsection{Projection for the High-Luminosity LHC}

\begin{figure}[htbp]
\centering
\includegraphics[width=0.6\textwidth]{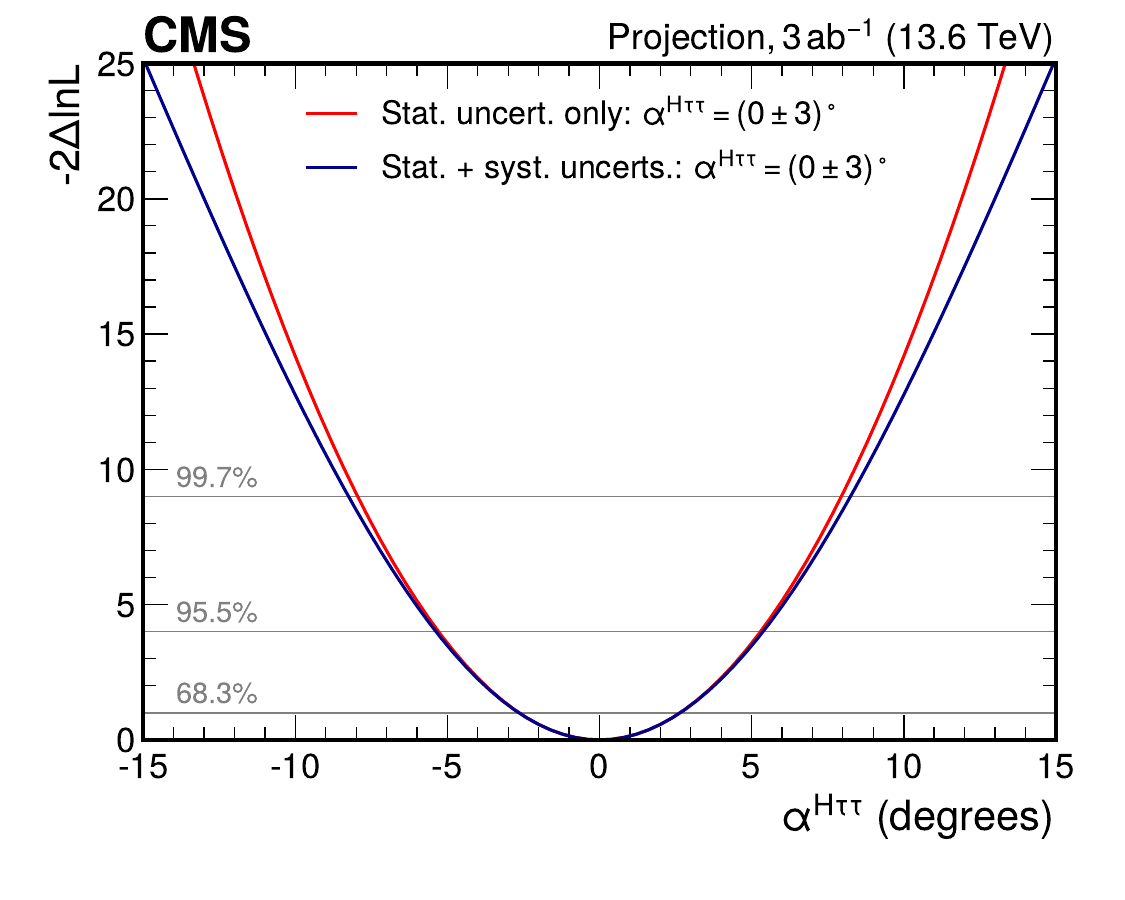}
\caption{Projections of the expected likelihood scans to an integrated luminosity of 3\abinv under two systematic uncertainty scenarios: all systematic uncertainties (including the bin-by-bin component) kept constant with respect to the 13.6\TeV analysis (blue) and no systematic uncertainties (red).
The expected precision on \alphacp is $3^\circ$ in both scenarios.}
\label{fig:extrapolation}
\end{figure}

A projection of the measurement of \alphacp to the end of the High-Luminosity LHC has been derived by extrapolating the results obtained at 13.6\TeV to an integrated luminosity of 3\abinv.
The extrapolation assumes that the analysis strategy and the performance of the CMS detector remain unchanged with respect to the 13.6\TeV analysis.
Two systematic uncertainty scenarios are considered: (i) all systematic uncertainties (including the bin-by-bin component) are kept constant with respect to the 13.6\TeV analysis, (ii) no systematic uncertainties are included.
The expected likelihood scans for these two scenarios are shown in Fig.~\ref{fig:extrapolation}. The expected precision on \alphacp is found to be $3^\circ$ in both cases. 
The systematic uncertainties typically affect the scan in regions away from the best fit point ($-2\Delta\ln \mathrm{L} \gtrsim 9$). 
The reduction in the overall uncertainty follows from the increased luminosity, which reduces the statistical uncertainty and, by enabling tighter constraints in the fit, also reduces the impact of the systematic uncertainties. 
We note that the expected precision on \alphacp exceeds that estimated for future $\Pe^{+}\Pe^{-}$ colliders, which is expected to be around $6^\circ$~\cite{Gritsan:2022php}.

\section{Summary}\label{sec:summary}

A measurement of the effective \CP mixing angle \alphacp between scalar and pseudoscalar \HTT couplings has been presented, 
using proton-proton collision data at $\sqrt{s}=13.6\TeV$ recorded with the CMS detector in 2022 and 2023, corresponding to an integrated luminosity of 62.4\fbinv. 
The mixing angle is measured to be $\alphacp = (36^{+33}_{-30}){}^\circ$, compared with an expected value of $(0 \pm 19)^\circ$. 
Despite using less than half the volume of data, the expected sensitivity is improved with respect to the previous CMS measurement that used data collected at $\sqrt{s}=13\TeV$, corresponding to an integrated luminosity of 138\fbinv.
The combined measurement using the $\sqrt{s}=13\TeV$ and $\sqrt{s}=13.6\TeV$ data sets yields $\alphacp = (7 \pm 16)^\circ$, with an expected value of $(0 \pm 14)^\circ$. 
This represents the most precise measurement by CMS of the \CP nature of the Higgs boson coupling to tau leptons, with an expected precision that is the best achieved by any experiment to date.
Finally, a projection is made for the expected precision of the measurement at the end of the High-Luminosity LHC, and is found to be approximately $3^\circ$.

\begin{acknowledgments}
We congratulate our colleagues in the CERN accelerator departments for the excellent performance of the LHC and thank the technical and administrative staffs at CERN and at other CMS institutes for their contributions to the success of the CMS effort. In addition, we gratefully acknowledge the computing centres and personnel of the Worldwide LHC Computing Grid and other centres for delivering so effectively the computing infrastructure essential to our analyses. Finally, we acknowledge the enduring support for the construction and operation of the LHC, the CMS detector, and the supporting computing infrastructure provided by the following funding agencies: SC (Armenia), BMBWF and FWF (Austria); FNRS and FWO (Belgium); CNPq, CAPES, FAPERJ, FAPERGS, and FAPESP (Brazil); MES and BNSF (Bulgaria); CERN; CAS, MoST, and NSFC (China); MINCIENCIAS (Colombia); MSES and CSF (Croatia); RIF (Cyprus); SENESCYT (Ecuador); ERC PRG and PSG, TARISTU24-TK10 and MoER TK202 (Estonia); Academy of Finland, MEC, and HIP (Finland); CEA and CNRS/IN2P3 (France); SRNSF (Georgia); BMFTR, DFG, and HGF (Germany); GSRI (Greece); MATE and NKFIH (Hungary); DAE and DST (India); IPM (Iran); SFI (Ireland); INFN (Italy); MSIT and NRF (Republic of Korea); MES (Latvia); LMTLT (Lithuania); MOE and UM (Malaysia); BUAP, CINVESTAV, CONACYT, LNS, SEP, and UASLP-FAI (Mexico); MOS (Montenegro); MBIE (New Zealand); PAEC (Pakistan); MSHE, NSC, and NAWA (Poland); FCT (Portugal); MESTD (Serbia); MICIU/AEI and PCTI (Spain); MOSTR (Sri Lanka); Swiss Funding Agencies (Switzerland); MST (Taipei); MHESI (Thailand); TUBITAK and TENMAK (T\"{u}rkiye); NASU (Ukraine); STFC (United Kingdom); DOE and NSF (USA).

\begin{sloppy}
\setlength\emergencystretch{\hsize}
\hyphenation{Rachada-pisek} Individuals have received support from the Marie-Curie programme and the European Research Council and Horizon 2020 Grant, contract Nos.\ 675440, 724704, 752730, 758316, 765710, 824093, 101115353, 101002207, 101001205, and COST Action CA16108 (European Union); the Leventis Foundation; the Alfred P.\ Sloan Foundation; the Alexander von Humboldt Foundation; the Science Committee, project no. 22rl-037 (Armenia); the Fonds pour la Formation \`a la Recherche dans l'Industrie et dans l'Agriculture (FRIA) and Fonds voor Wetenschappelijk Onderzoek contract No. 1228724N (Belgium); the Beijing Municipal Science \& Technology Commission, No. Z191100007219010, the Fundamental Research Funds for the Central Universities, the Ministry of Science and Technology of China under Grant No. 2023YFA1605804, the Natural Science Foundation of China under Grant No. 12535004, and USTC Research Funds of the Double First-Class Initiative No.\ YD2030002017 (China); the Ministry of Education, Youth and Sports (MEYS) of the Czech Republic; Agence Nationale de la Recherche grant ANR-22-CE31-0002 and the Interdisciplinary Thematic Institute in Quantum Science and Nanomaterials (QMat) (France); the Shota Rustaveli National Science Foundation (Georgia); the Deutsche Forschungsgemeinschaft (DFG), among others, under Germany's Excellence Strategy -- EXC 2121 ``Quantum Universe" -- 390833306, and under project number 400140256 - GRK2497; the Hellenic Foundation for Research and Innovation (HFRI), Project Number 2288 (Greece); the Hungarian Academy of Sciences, the New National Excellence Program - \'UNKP, the NKFIH research grants K 131991, K 138136, K 143460, K 143477, K 147557, K 146913, K 146914, K 147048, TKP2021-NKTA-64, and 2025-1.1.5-NEMZ\_KI-2025-00004, and MATE KKP and KKPCs Research Excellence and Flagship Research Groups grants (Hungary); the Council of Science and Industrial Research, India; ICSC -- National Research Centre for High Performance Computing, Big Data and Quantum Computing, FAIR -- Future Artificial Intelligence Research, and CUP I53D23001070006 (Mission 4 Component 1), funded by the NextGenerationEU program, the Italian Ministry of University and Research (MUR) under Bando PRIN 2022 -- CUP I53C24002390006, PRIN PRIMULA 2022RBYK7T (Italy); the Latvian Council of Science; the Ministry of Science and Higher Education, project no. 2022/WK/14, and the National Science Centre, contracts Opus 2021/41/B/ST2/01369, 2021/43/B/ST2/01552, 2023/49/B/ST2/03273, and the NAWA contract BPN/PPO/2021/1/00011 (Poland); the Funda\c{c}\~ao para a Ci\^encia e a Tecnologia (Portugal); the National Priorities Research Program by Qatar National Research Fund; MICIU/AEI/10.13039/501100011033, ERDF/EU, ``European Union NextGenerationEU/PRTR", projects PID2022-142604OB-C21, PID2022-139519OB-C21, PID2023-147706NB-I00, PID2023-148896NB-I00, PID2023-146983NB-I00, PID2023-147115NB-I00, PID2023-148418NB-C41, PID2023-148418NB-C42, PID2023-148418NB-C43, PID2023-148418NB-C44, PID2024-158190NB-C22, RYC2021-033305-I, RYC2024-048719-I, CNS2023-144781, CNS2024-154769 and Plan de Ciencia, Tecnolog{\'i}a e Innovaci{\'o}n de Asturias, Spain; the Chulalongkorn Academic into Its 2nd Century Project Advancement Project, the National Science, Research and Innovation Fund program IND\_FF\_68\_369\_2300\_097, and the Program Management Unit for Human Resources \& Institutional Development, Research and Innovation, grant B39G680009 (Thailand); the Eric \& Wendy Schmidt Fund for Strategic Innovation through the CERN Next Generation Triggers project under grant agreement number SIF-2023-004; the Kavli Foundation; the Nvidia Corporation; the SuperMicro Corporation; the Welch Foundation, contract C-1845; and the Weston Havens Foundation (USA).
\end{sloppy}
\end{acknowledgments}\section*{Data availability} Release and preservation of data used by the CMS Collaboration as the basis for publications is guided by the  \href{https://doi.org/10.7483/OPENDATA.CMS.1BNU.8V1W}{CMS data preservation, re-use and open access policy}.

\bibliography{auto_generated}
\cleardoublepage \appendix\section{The CMS Collaboration \label{app:collab}}\begin{sloppypar}\hyphenpenalty=5000\widowpenalty=500\clubpenalty=5000\cmsinstitute{Yerevan Physics Institute, Yerevan, Armenia}
{\tolerance=6000
A.~Gevorgyan\cmsorcid{0000-0003-2751-9489}, A.~Hayrapetyan, V.~Makarenko\cmsorcid{0000-0002-8406-8605}, A.~Tumasyan\cmsAuthorMark{1}\cmsorcid{0009-0000-0684-6742}
\par}
\cmsinstitute{Marietta Blau Institute for Particle Physics, Vienna, Austria}
{\tolerance=6000
W.~Adam\cmsorcid{0000-0001-9099-4341}, L.~Benato\cmsorcid{0000-0001-5135-7489}, T.~Bergauer\cmsorcid{0000-0002-5786-0293}, M.~Dragicevic\cmsorcid{0000-0003-1967-6783}, S.~Gundacker\cmsorcid{0000-0003-2087-3266}, A.K.~Guven\cmsorcid{0009-0004-5670-5138}, P.S.~Hussain\cmsorcid{0000-0002-4825-5278}, M.~Jeitler\cmsAuthorMark{2}\cmsorcid{0000-0002-5141-9560}, N.~Krammer\cmsorcid{0000-0002-0548-0985}, A.~Li\cmsorcid{0000-0002-4547-116X}, D.~Liko\cmsorcid{0000-0002-3380-473X}, M.~Matthewman, J.~Schieck\cmsAuthorMark{2}\cmsorcid{0000-0002-1058-8093}, R.~Sch\"{o}fbeck\cmsAuthorMark{2}\cmsorcid{0000-0002-2332-8784}, M.~Shooshtari\cmsorcid{0009-0004-8882-4887}, M.~Sonawane\cmsorcid{0000-0003-0510-7010}, N.~Van~Den~Bossche\cmsorcid{0000-0003-2973-4991}, W.~Waltenberger\cmsorcid{0000-0002-6215-7228}, C.-E.~Wulz\cmsAuthorMark{2}\cmsorcid{0000-0001-9226-5812}
% ...
\par}
\vskip\cmsinstskip
\dag:~Deceased\\
$^{1}$Also at Yerevan State University, Yerevan, Armenia\\
$^{2}$Also at TU Wien, Vienna, Austria\\
% ...
\end{sloppypar}
\end{document}